\documentclass[12pt,preprint]{aastex} 

\newcommand{\gapprox}{$\stackrel {>}{_{\sim}}$}   
\newcommand{\lapprox}{$\stackrel {<}{_{\sim}}$}

\slugcomment{To appear in Ap.J.}

\begin{document}

\title{Near Infrared Spectroscopic Monitoring of EXor variables:\\ First Results
\thanks{Based on observations collected at the AZT-24 telescope
(Campo Imperatore, Italy)}}

\author{
   D.Lorenzetti\altaffilmark{2}
   V.M.Larionov\altaffilmark{3,4},
   T.Giannini\altaffilmark{2},
   A.A.Arkharov\altaffilmark{4},
   S.Antoniucci\altaffilmark{2},
   B.Nisini\altaffilmark{2},
          and
    A.Di Paola\altaffilmark{2}
}
%
\altaffiltext{2}{INAF - Osservatorio Astronomico di Roma, via
Frascati 33,I-00040  Monte Porzio, Italy, dloren, giannini, antoniucci,
nisini, dipaola, dloren@oa-roma.inaf.it}
\altaffiltext{3}{Astronomical Institute of St.Petersburg
University, Russia, vlar@nm.ru}
\altaffiltext{4}{Central Astronomical Observatory of Pulkovo,
Pulkovskoe shosse 65, 196140 St.Petersburg,
Russia, arkharov@mail.ru}
%
\date{Received;Accepted}


\begin{abstract}
We present low resolution ($\mathcal{R}$ $\sim$ 250) spectroscopy
in the near-IR (0.8 to 2.5 $\mu$m) of the EXor variables. These
are the initial results (obtained during the period 2007-2008)
from a long term photometric and spectroscopic program aimed to
study the variability in the accretion processes of pre-Main
Sequence (PMS) stars, by correlating the continuum fluctuations
with the spectroscopical properties.
Eight sources have been observed
in different epochs, for a total of 25 acquired spectra. EXor
spectra show a wide variety of emission features dominated by HI
recombination (Paschen and Brackett series). We have investigated
whether line and continuum variability could be due to a variable
extinction, but such hypothesis is applicable only to the peculiar
source PV Cep. By comparing the observed spectra with a wind
model, mass loss rates in the range (2-10) 10$^{-8}$ M$_{\sun}$
yr$^{-1}$ are derived, along with other wind parameters.
Consistent results are also obtained by assuming that HI lines are
due to accretion. CO overtone is also detected in the majority of
the sources both in absorption and in emission. It appears to come
from regions more compact than winds, likely the stellar
photosphere (when in absorption) and the circumstellar disk (when
in emission). NaI and CaI IR lines behave as the CO does, thus
they are thought to arise in the same locations. For some targets
multiple spectra correspond to different activity stages of the
source. Those exhibiting the largest continuum variation at 2
$\mu$m ($\Delta$K \gapprox  1 mag) present a significant line flux
fading during the continuum declining phases. In particular, CO
absorption (emission) appears associated to inactive (active)
stages, respectively.
\end{abstract}




%

\section{Introduction}

EXor stars, originally defined by Herbig (1989), are Pre-Main
Sequence objects, with ages of about 10$^6$ years, characterized
by intense (3-4 magnitudes) and short living (months, one year)
outbursts superposed to longer (some years) quiescence periods.
According to a widely accepted picture, such objects are accreting
material from a circumstellar disk, through rapid and intermittent
accretion events that generate thermal instabilities in the disk
itself and, eventually, outbursts phenomena (Hartmann \& Kenyon
1985). Indeed the accreted matter migrates toward the central star
and it is suddenly halted where the inner disk is truncated (at
few stellar radii) by the dipolar stellar magnetic field; then it
is channelled onto the stellar surface along the magnetic field
lines (e.g. Shu et al. 1994). When such material violently falls
onto the stellar surface produces a shock that cools by emitting a
hot continuum, often called veiling. As a consequence of the
accretion event, strong winds (in some cases also collimated jets)
emerge from the rotating star/disk system. While the accretion
phenomenology is difficult to be directly observed, the
observations of young objects more likely reveal the presence of
outflowing matter whose rate is often exploited to quantitatively
determine the rate of the infalling one, by invoking the rough
proportion $\dot{M}_{wind}/\dot{M}_{acc}$ \lapprox~ 0.1 (Shu et
al. 2000; K\"{o}nigl \& Pudritz 2000).\\

The interaction with a close binary companion is also invoked as
an alternative mechanism to produce accretion disk instabilities
and consequent outbursts (Clarke, Lin \& Pringle 1990; Bonnell \&
Bastien 1992). Indeed, the majority of EXor do have a close
companion which could trigger the flares when it passes at the
periastron. A suggestion in this sense has been provided to
account for the rapid variations of UZ Tau E (Jensen et al. 2007),
and an increasing number of observational studies are currently
oriented in such direction. However, extending the close companion
interpretation to the whole class has not proved to be able to
explain the complex phenomenology (e.g. the timescale variability)
observed in EXor's. The present paper focusses on the correlations
between continuum and lines variability, be they intrinsic to the
source or triggered by the external environment. Continuum and
line observations at different frequencies trace phenomena that
occur in completely different regions of these complex systems
(from the outer disk to the stellar surface and even in the
external envelope and in the chromosphere). Although the
flaring-up events represent a very important phase of the Pre-Main
Sequence life, the continuum and spectral line variability are
rarely correlated and very few spectroscopic studies exist to date
that compare outburst vs quiescence observational properties.
Herbig (2007) provides an optical monitoring of EX Lup (the
prototype of the EXor class) which covers a long period during
which four subsequent flareups occurred. The spectroscopic
consequences of the outburst are the appearance of both a hot
continuum and an emission line structure where inverse P Cyg
absorption features are superposed. Five EXor candidates (NY Ori,
V1118 Ori, V1134 Ori, V1184 Tau, and V350 Cep) have been compared
both photometrically and spectroscopically with EX Lup, looking
for common features (Herbig 2008). Spectroscopical monitoring
studies in the visual band have been presented for other EXor's,
namely DR Tau (Beristain et al. 1998), VY Tau (Herbig 1990), PV
Cep (Cohen et al. 1981). Typically, when the star is inactive the
spectrum shows absorption features  of a M0 star, while a
remarkable emission spectrum appears during the active phases.
Similar studies in the IR band are still missing, although they
should be crucial during both inactive and active phases: in the
former stages they allow us to investigate how the properties of
the circumstellar matter prior the outburst will influence,
through the accretion, the outburst itself; while, in the latter
stages the IR spectroscopy is a suitable tool to sample how the
circumstellar material is altered by intermittent mass loss. In
this wavelength range, the only monitored variation is a decrease
of the line emission intensity by more than a factor of six
detected in the near IR spectrum of the EXor V1118 Ori, passing
from active to inactive status (Lorenzetti et al. 2007, hereinafter Paper II).\\

Quite recently different outbursting and fading phases of another
young object, i.e. V1647 Ori, have been spectroscopically
monitored in IR (Gibb et al. 2006; Acosta-Pulido et al. 2007)
demonstrating how the post-outburst phase is characterized by a
declining temperature of the hot CO gas formed in the inner part
of the disk, and by a substantial decrease of the fast wind. In
particular, after one year of quiescence, HI and He recombination
are decreased by one order of magnitude and CO lines appear in
absorption instead of in emission (Aspin, Beck \& Reipurth 2008).
These latter authors argue in favour of the membership of this
embedded object to the EXor class (instead to the FUor one), since
its recent activity lasted only two years and 40 years ago it was
observed in outburst. Near-IR spectra have been provided for the
embedded variable OO Ser (Hodapp 1999; K\'{o}sp\'{a}l et al.
2007), but, again, some doubts on its attribution to the EXor
class have been risen by the authors themselves by invoking both
the wavelength-independence of the fading and the variability
timescales. A somehow unified interpretative scheme able to
account for the observed near-IR spectra of EXor, does not exist,
at the moment. To understand the correlation between the gas
properties due to accretion/ejection activity and the continuum
variation in the near-IR, we have started a spectroscopical
monitoring program in the range 0.8-2.5 $\mu$m, and the first
results are presented here. Our scope is also to ascertain whether
or not the conclusions mentioned above for individual objects can
be ascribed to the whole EXor class and to examine the
similarities vs. differences that characterize their near-IR
spectra.

In Paper II the IR (1-100 $\mu$m) photometric properties of the
EXor have been discussed, while in the present paper more emphasis
is given to the spectroscopic properties and to their correlations
with the continuum status. For some targets, those presented here
represent the first IR spectra ever taken, hence this paper aims
hopefully to contribute in creating an initial data-base for
future comparison purposes. After having defined the EXors sample
and provided the details of our observations in Sect.2, we present
in Sect.3 our spectroscopic results. In Sect.4 these are
interpreted by means of a wind model to derive the physical
properties of the emitting gas; our concluding remarks are given
in Sect.5.

\section{Observations}

The list of the investigated targets is given in
Table~\ref{parameters:tab} along with some relevant parameters.
The tenth column indicates whether or not a close companion has
been found, providing in the affirmative cases the inter-distance
in arcsec. Our list is essentially the one originally provided by
Herbig (1989), where EX Lup is missing because of its southern
location unaccessible by our instrumentation.

Near-IR data were obtained at the 1.1m AZT-24 telescope located at
Campo Imperatore (L'Aquila - Italy) equipped with the
imager/spectrometer SWIRCAM (D'Alessio et al. 2000), which is
based on a 256$\times$256 HgCdTe PICNIC array. Low resolution
($\mathcal{R}$ $\sim$ 250) spectroscopy is obtained by means of
two IR grisms G$_{blue}$ and G$_{red}$ covering the ZJ (0.83 -
1.34 $\mu$m) and HK (1.44 - 2.35 $\mu$m) bands, respectively, in
two subsequent exposures. The long slit is not orientable in
position angle, and it samples a pre-defined portion of the focal
plane, 2$\times$260 arcsec$^2$ in the east-west direction.

Long-slit spectroscopy was carried out in the standard
ABB$\arcmin$A$\arcmin$ mode with a total integration time ranging
between 800 and 1200 sec. In Table~\ref{log:tab} the Log of our
observations is given. The spectral images were flat-fielded,
sky-subtracted, and corrected for the optical distortion in both
the spatial and spectral directions. The object spectra were corrected for
the atmospheric spectral response dividing them by the spectrum of a
telluric O-type star, having normalized this latter for the black-body
spectrum at the stellar temperature and replaced its intrinsic hydrogen
absorption lines with the black-body function at the same wavelengths.
Wavelength calibration was derived from the OH lines
present in the raw spectral images, while flux calibration was
obtained from our photometric data taken to implement our database
on IR observations of EXors.\\
Such photometry, carried out on the same night of the
spectroscopy, is given in Table~\ref{log:tab} in form of J,H and K
magnitudes, provides an indication of the current brightness
status of the source compared with the historical behavior derived
from the literature (see Paper II). The same Table~\ref{log:tab}
is complemented with information about the spectroscopic
observations in the near IR (1-2.5 $\mu$m) available in the
literature.

\section{Results}

\begin{figure}
 \centering
   \includegraphics [width=15 cm] {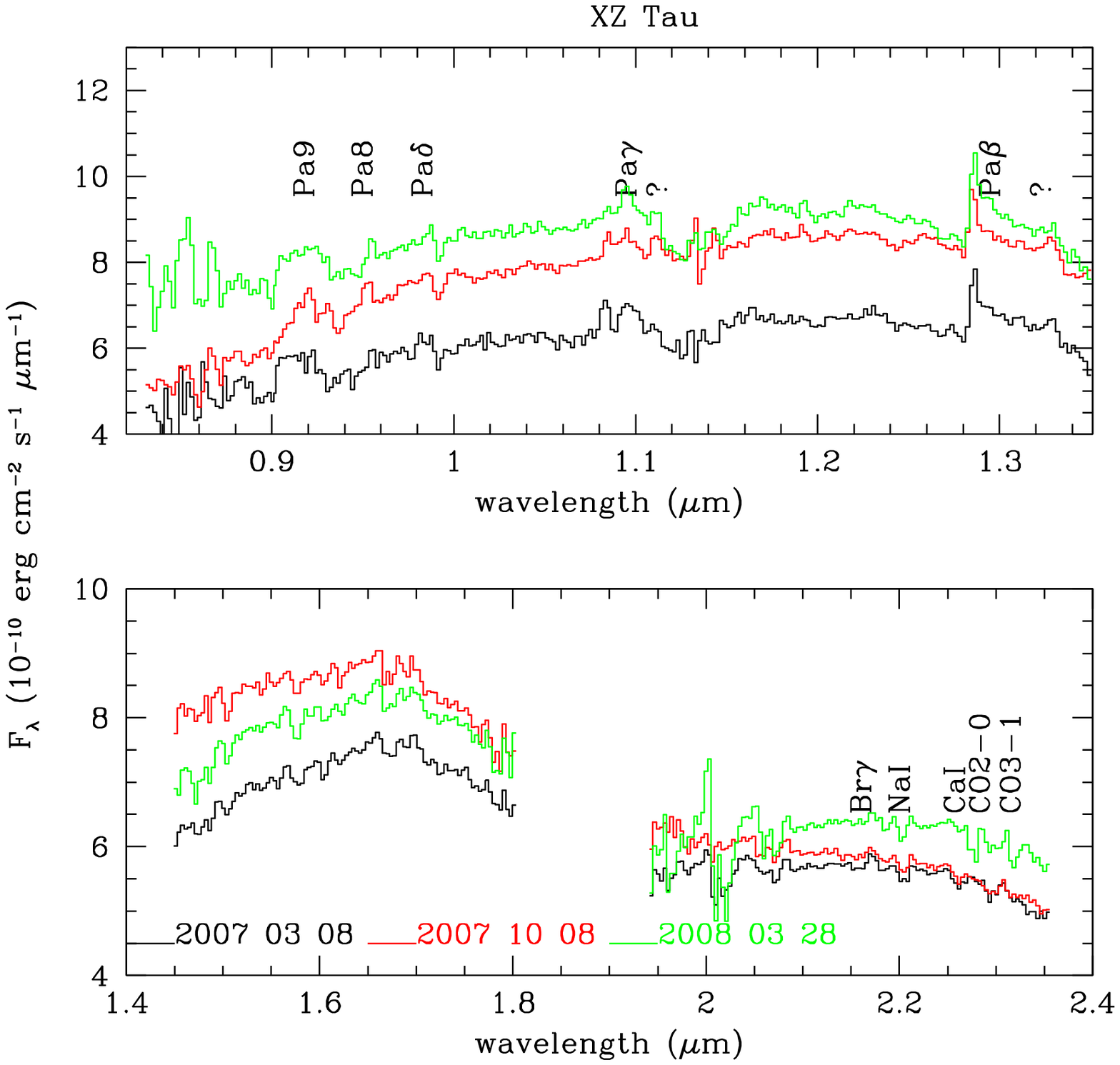}
   \caption{Near IR spectrum of XZ Tau. Detected line are identified and listed in Table~\ref{linesxztau:tab}.
   \label{XZTauspectrum:fig}}
\end{figure}

\begin{figure}
 \centering
   \includegraphics [width=15 cm] {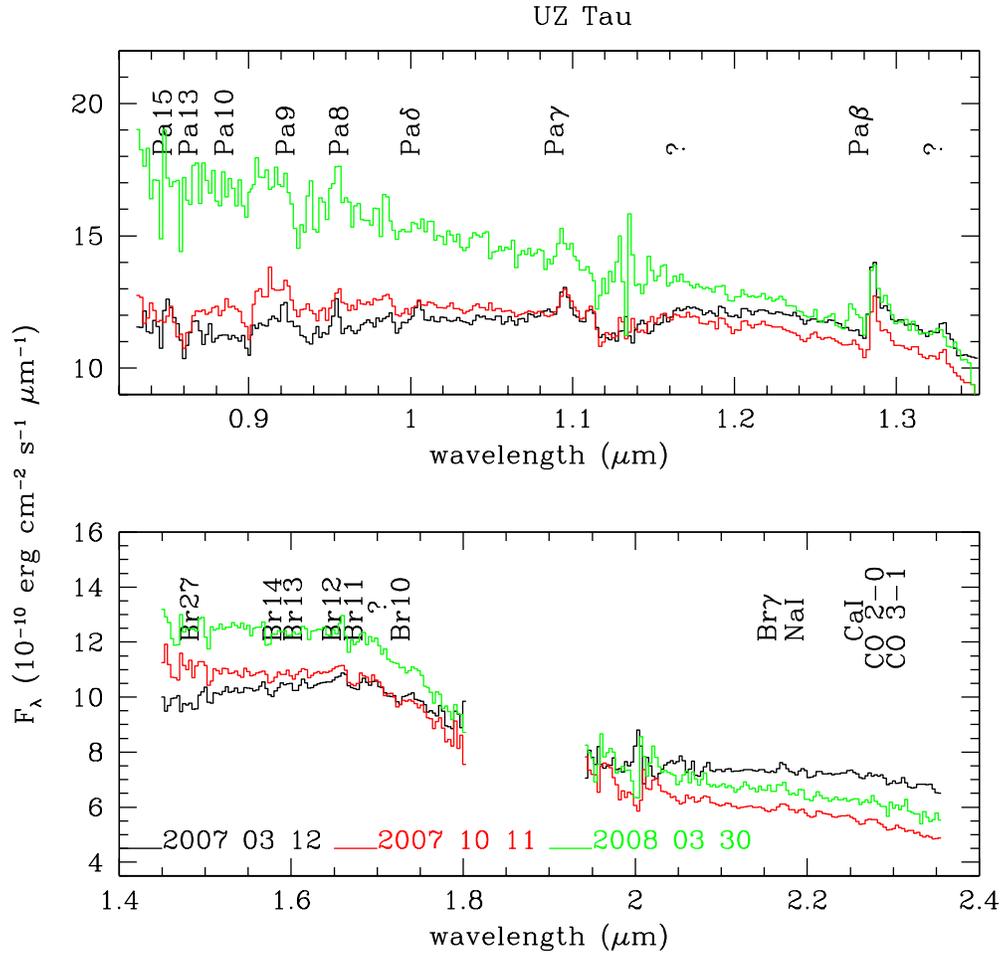}
   \caption{Near IR spectrum of UZ Tau. Detected line are identified and listed in Table~\ref{linesuztau:tab}.
   E and W components are not resolved in the spectral mode.
   \label{UZTauspectrum:fig}}
\end{figure}

\begin{figure}
 \centering
   \includegraphics [width=15 cm] {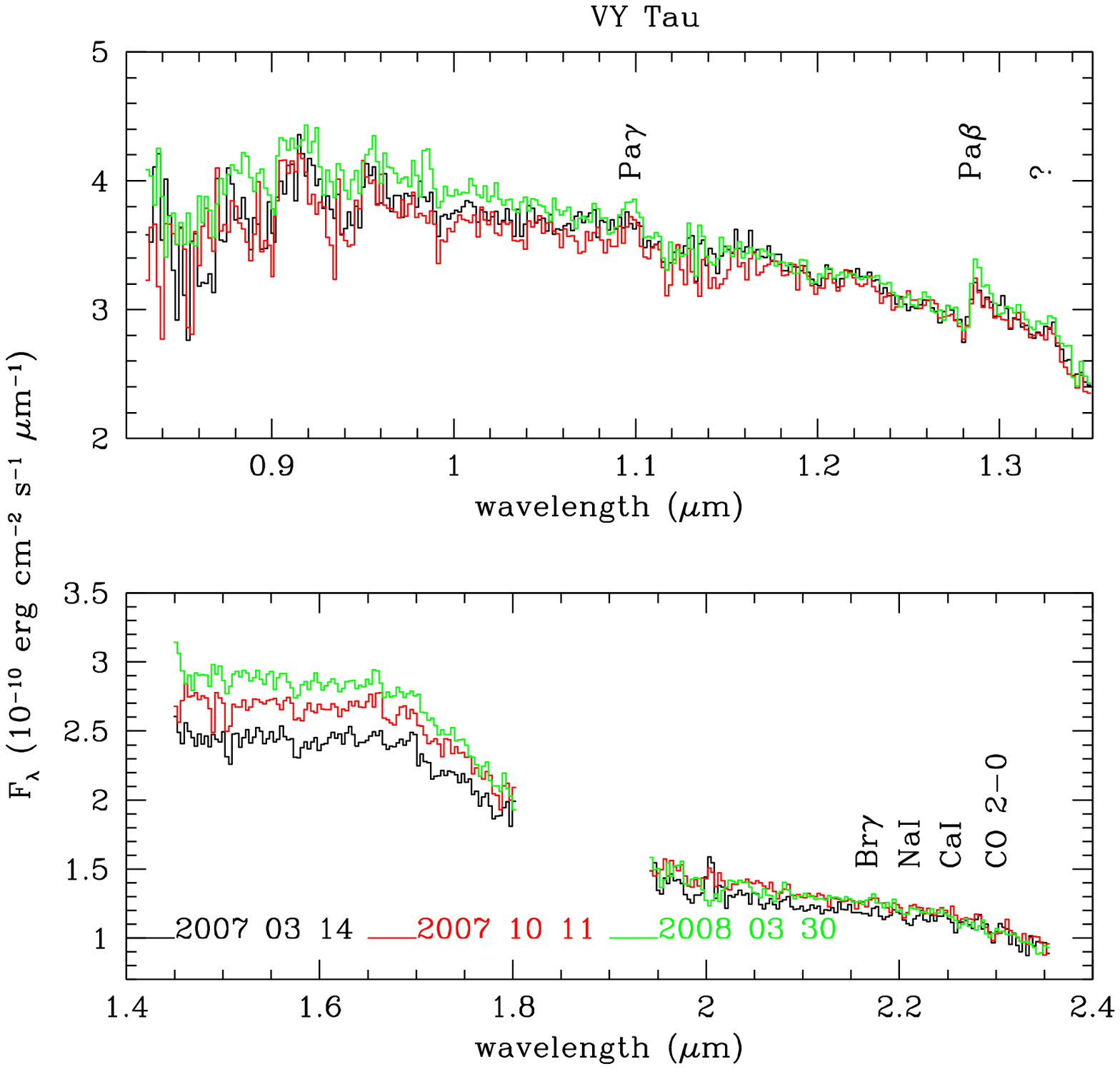}
   \caption{Near IR spectrum of VY Tau. Detected line are identified and listed in Table~\ref{linesvytau:tab}.
   \label{VYTauspectrum:fig}}
\end{figure}

\begin{figure}
 \centering
   \includegraphics [width=15 cm] {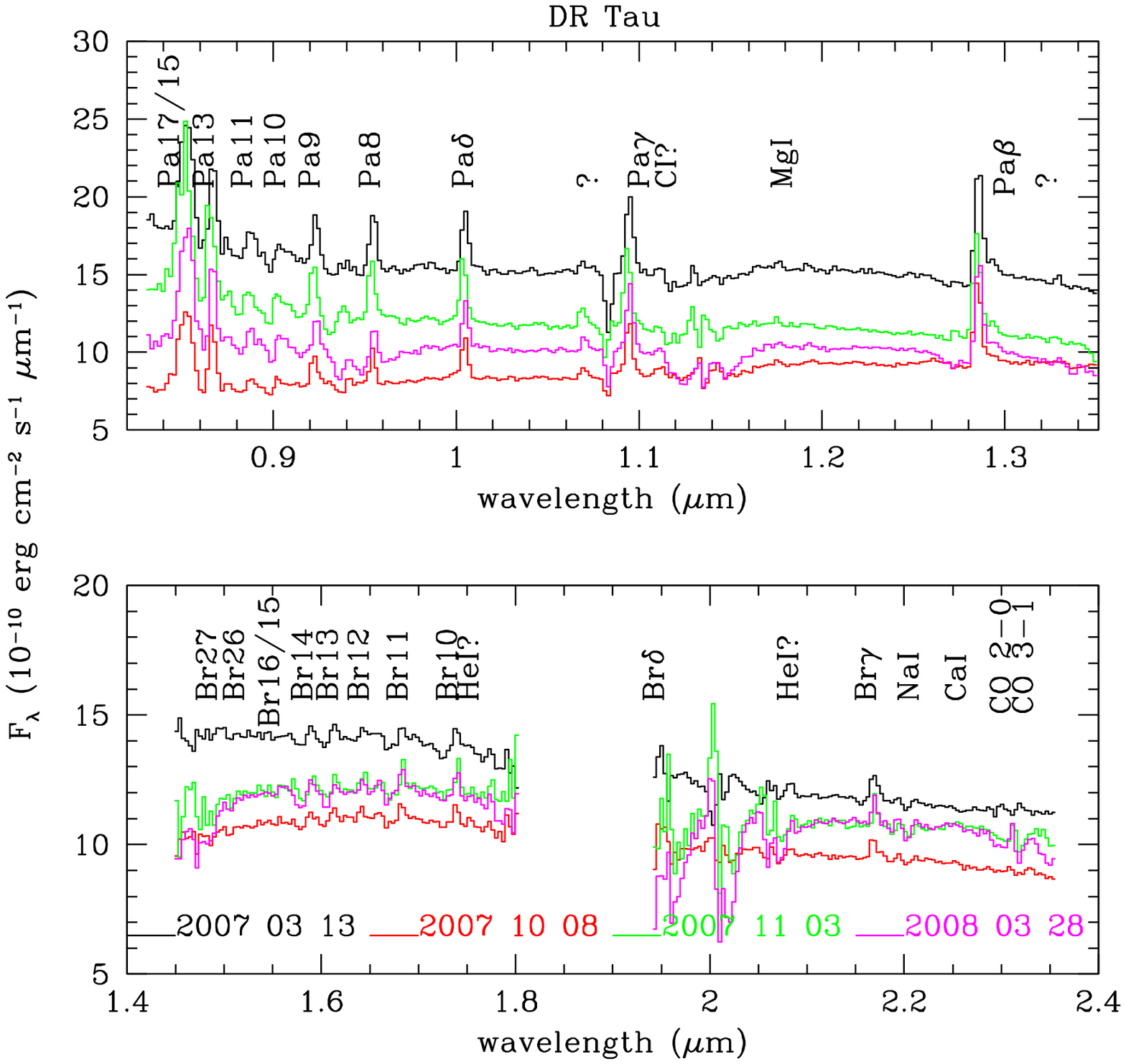}
   \caption{Near IR spectrum of DR Tau. Detected line are identified and listed in Table~\ref{linesdrtau:tab}.
   \label{DRTauspectrum:fig}}
\end{figure}

\begin{figure}
 \centering
   \includegraphics [width=15 cm] {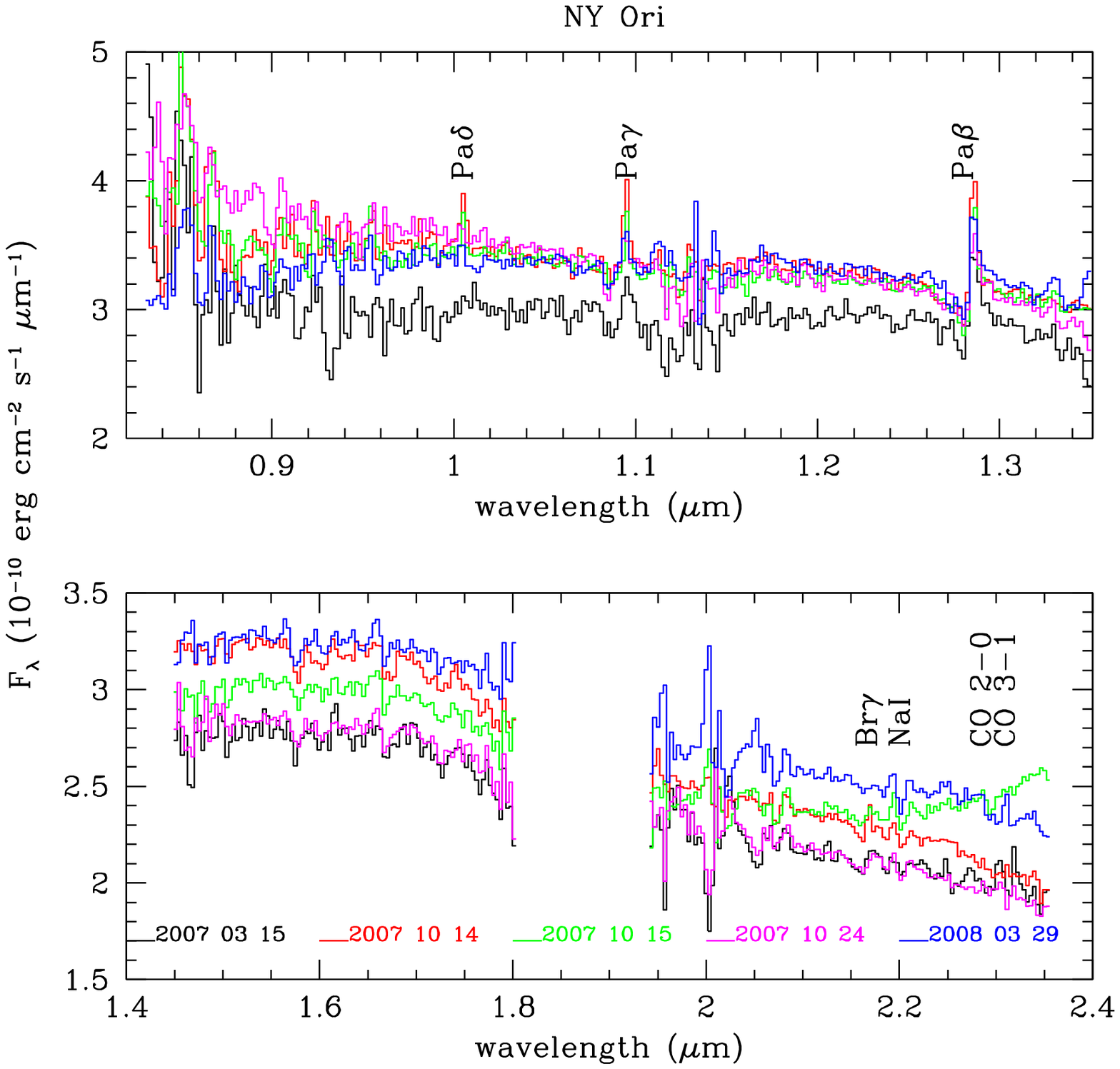}
   \caption{Near IR spectrum of NY Ori. Detected line are identified and listed in Table~\ref{linesnyorise:tab}.
   \label{NYOri_SEspectrum:fig}}
\end{figure}

\begin{figure}
 \centering
   \includegraphics [width=15 cm] {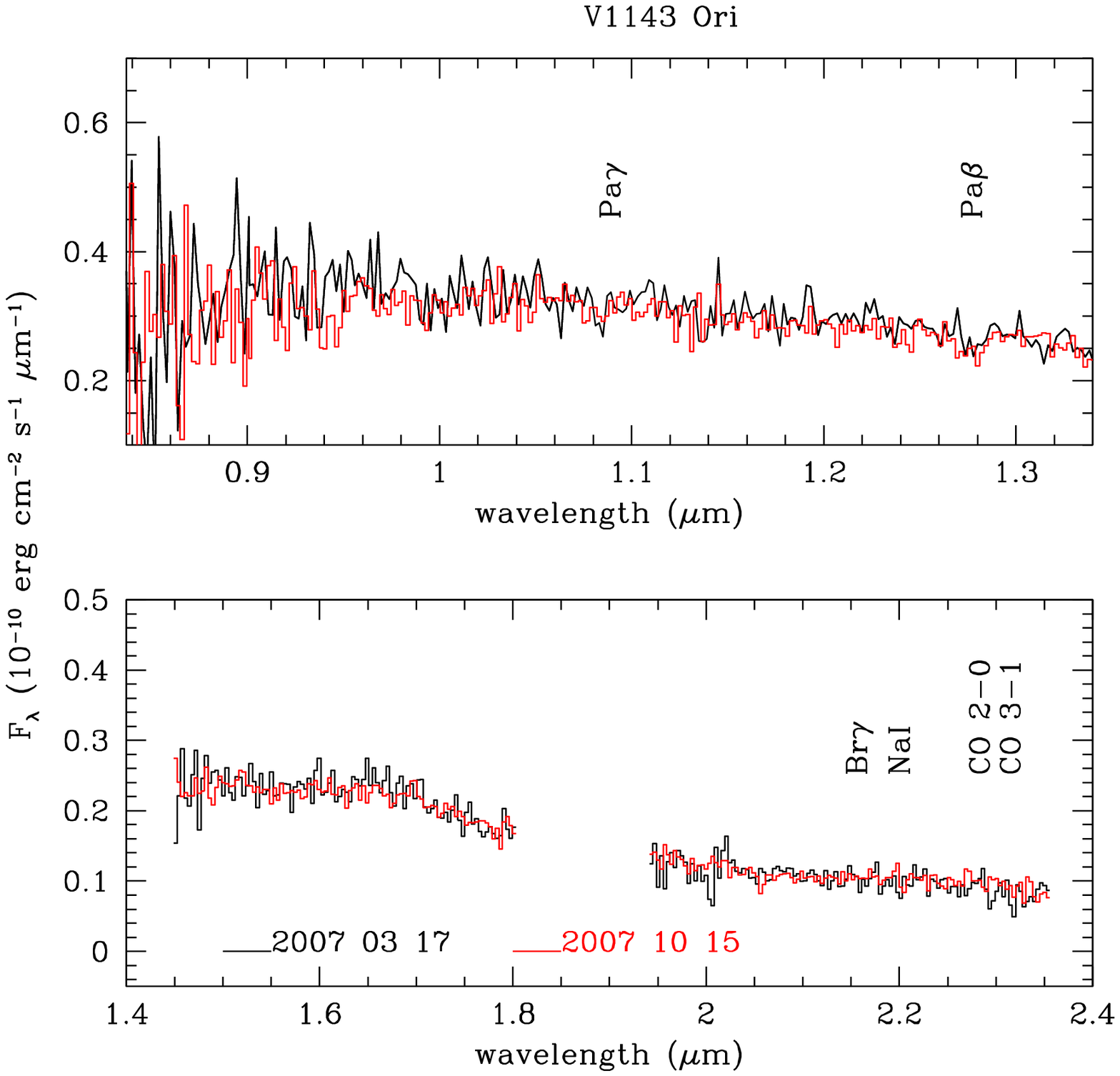}
   \caption{Near IR spectrum of V1143 Ori. Detected line are identified and listed in Table~\ref{linesv1143ori:tab}.
   \label{V1143Orispectrum:fig}}
\end{figure}

\begin{figure}
 \centering
   \includegraphics [width=15 cm] {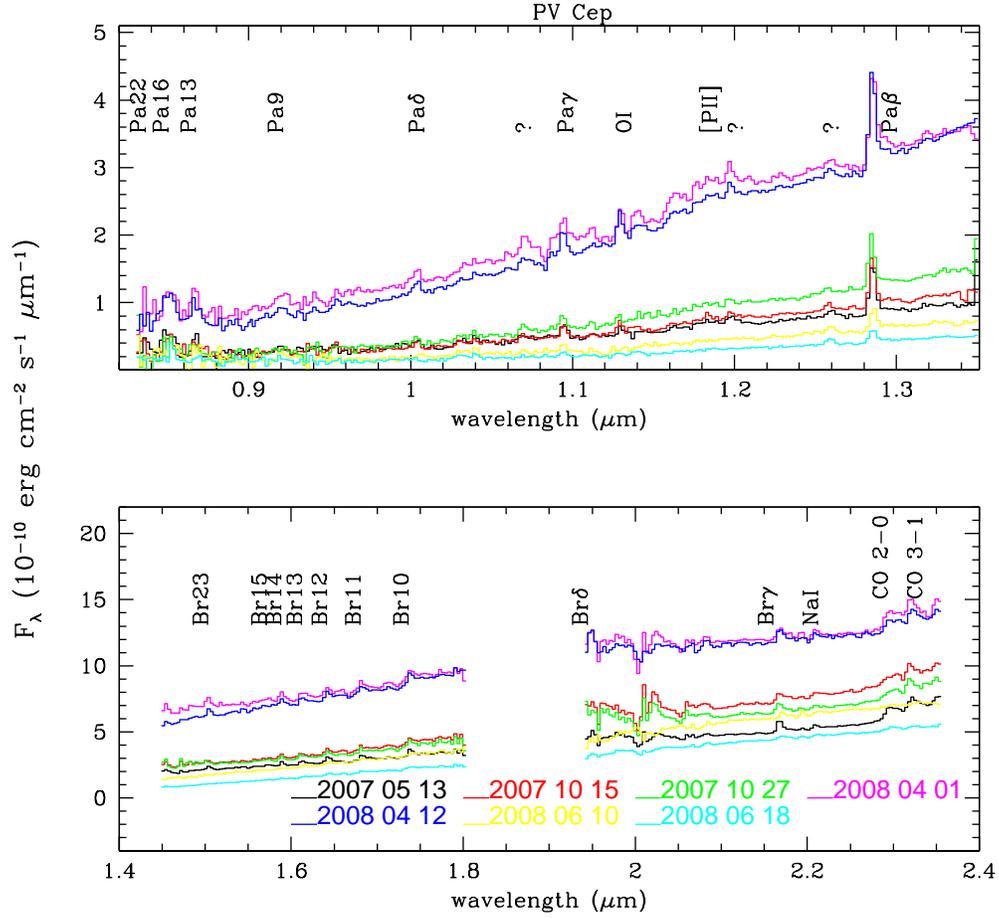}
   \caption{Near IR spectrum of PV Cep. Detected line are identified and listed in Table~\ref{linespvcep:tab}.
   \label{PVCepspectrum:fig}}
\end{figure}

The results given here represent the first survey of EXors spectra
carried out in a systematic way, namely by observing a complete
sample with the same equipment and by adopting the same
observational modalities and reduction techniques. For two objects
(NY Ori and V1143 Ori) these represent the first near-IR spectra
ever obtained. In the following, the EXor source V1118 Ori  will
be incorporated in the discussion, although its spectroscopic data
have been already presented elsewhere (Papers I and II). The
calibrated spectra are given in Figures from
\ref{XZTauspectrum:fig} to \ref{PVCepspectrum:fig} and the derived
unreddened line fluxes in Tables from \ref{linesxztau:tab} to
\ref{linespvcep:tab}. The lines showing a S/N ratio between 2 and
3 are still listed in the Tables and marked with an asterisk, but
they are not considered in the following analysis. In the same
Tables, for any line flux the equivalent width (EW) is given, as
well. The EW does not represent a straightforward spectral
diagnostic, since it crucially depends on the continuum level;
however, since their values vary significantly, EW's are
signalling that spectroscopical variations do not merely follow
the continuum ones.

At our sensitivity
and at the epoch of our survey, some objects appear as emission
lines rich (UZ Tau, DR Tau, V1118 Ori and PV Cep), other (XZ Tau,
VY Tau, NY Ori and V1143 Ori) present just few features, usually
in emission: these are plausibly the brightest lines of a spectrum
weaker, but intrinsically composed by the same features present in
the line rich objects. These latter spectra are dominated by
hydrogen recombination (Brackett and Paschen series) which signals
the presence of ionized gas close to the star. In all the spectra
Pa$\beta$ is detected as a broad feature at the corresponding
wavelength: at $\mathcal{R}$ $\sim$ 250, this typical feature may
be due to the closeness of Pa$\beta$ and HeI at 1.285 $\mu$m,
however, because of the its relative faintness, the latter cannot
be confidently resolved.

The CO overtone features v=2-0, v=3-1 (both in emission and
absorption) are clearly detected in the majority of cases (7 out
of 8); the same occurs in several young stellar objects (Carr
1989), at variance with the majority of the FUor where CO bands
are revealed always in absorption (Hartmann, Hinkle \& Calvet
2004). The CO features are highly variable on relatively short
(some months) timescales. Besides CO flux variations, during our
monitoring period the CO bands have been observed even to change
from emission to absorption (DR Tau, V1118 Ori and NY Ori).
Such a behaviour is quite typical of low luminosity
young stellar objects, as monitored by Biscaya et al. (1997).

Atomic features of both NaI at 2.208 $\mu$m and, more rarely, CaI
at 2.267 $\mu$m, are also detected: in emission in those cases of
CO emission, and in absorption when also CO occurs in absorption.
This circumstance suggests the presence of a common origin for Na,
Ca and CO transitions; this topic will be discussed in Sect.4.5.
We note that a NaI line has been also detected, in absorption
(Herbig et al. 2001), in the near-IR spectrum of EX Lup, the
prototype of the EXor class. Molecular hydrogen and ionized iron
emission are absent at our sensitivity level in all the spectra
indicating that shocks are not a major excitation mechanism in
EXor environments. Indeed, weak shock excited features below our
sensitivity threshold have been detected in few cases (see
Table~\ref{log:tab}): in particular we mention here the
blue-shifted [FeII] emission lines (at 1.53 and 1.64 $\mu$m)
likely associated with the spatially resolved jet of PV Cep
(Hartmann et al. 1994).

By looking at the near IR continuum shape of our sources, a broad
water absorption feature near 1.9 $\mu$m seems recognizable in
some objects (mainly XZ Tau, UZ Tau and VY Tau). Some evidence
appears also in favour of a similar water feature near 1.4 $\mu$m,
which usually goes with the 1.9 $\mu$m one in very late low mass
stars and brown dwarfs (Lan\c{c}on \& Rocca-Volmerange 1992).
These three sources indeed present the latest spectral types among
the EXor sample and display also and other (NaI, CaI) photospheric
absorptions (see Sections 4.3 and 4.4), suggesting a similar
origin for the water bands, as well.

As a general trend, the observed spectra of the EXor are much more
similar to those of accreting T Tauri stars (Greene \& Lada 1996)
than the FUor ones. All these latter have spectra always dominated
by absorption lines (Reipurth \& Aspin 1997), apart a couple of
exceptions.

Spectroscopic variability is more or less evident in all the
monitored sources and is often accompanied by significant
variations of the line equivalent widths (see Tables from
\ref{linesxztau:tab} to \ref{linespvcep:tab}); this means that
spectroscopical variations are not a mere consequences of the
continuum ones, but they are related to the source brightness
through a less trivial link.

\section{Discussion}

Before analyzing in the next Sections the results of our near-IR spectral survey,
we shortly comment on the variability presented by the EXor near-IR spectra
collected so far in the literature.

The last 30 years of near-IR spectroscopic results of EXor are
summarized in Table~\ref{linevar:tab} along with the data of the
present survey given in boldface). Literature data are available
only for the indicated transitions and are given in form of
calibrated flux densities or equivalent width (EW), accordingly to
the original papers. This twofold way of presenting the data
forces us to examine line flux and EW variability of a given
source as two separate sets. A first glance at
Table~\ref{linevar:tab} indicates that only 14 spectra in total
were known prior of our survey, hence it is clearly evident that
studying IR line variability of EXor has not represented so far a
major interest, despite of the significant spectral changes
detected at optical wavelengths that, however, sample inner and
warmer regions. The past few and sparse observations indicate that
some IR line variation (by about a factor of 2) was recognizable
but on very uncertain timescales. Taking into account also our
data, we can estimate an IR line variability from a factor of 2 up
to an order of magnitude and typical timescales from months to
years, respectively.

\subsection{Spectroscopic vs. photometric variability}

\begin{figure}
 \centering
   \includegraphics [width=15 cm] {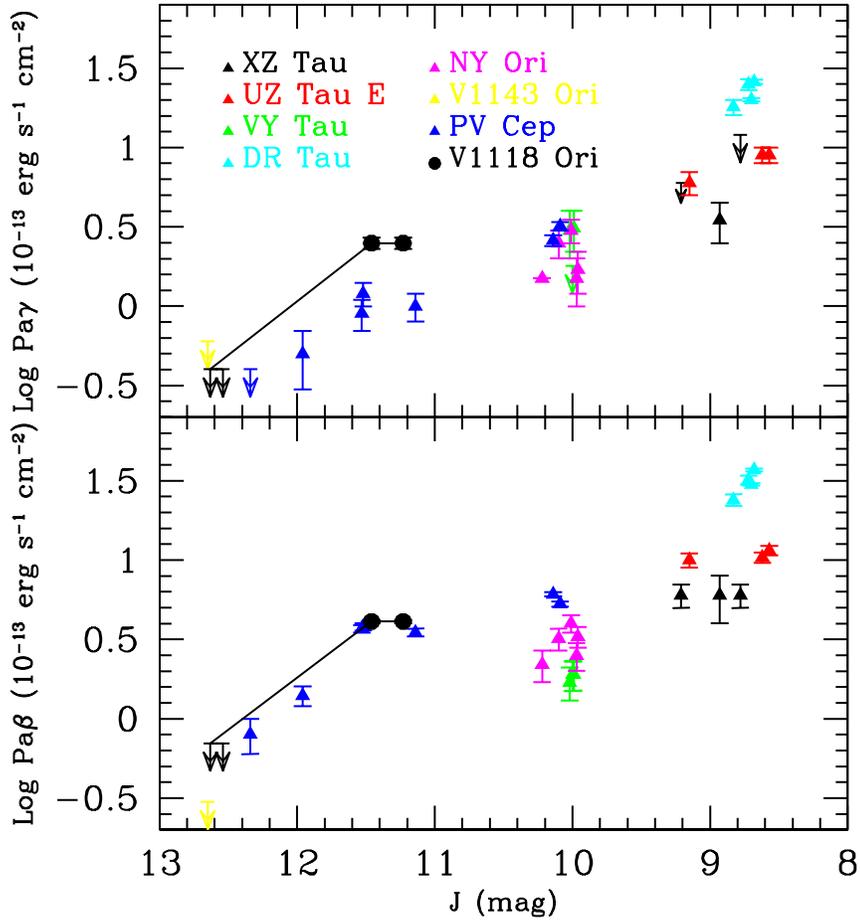}
   \caption{Observed Pa$\gamma$ and P$\beta$ fluxes as a function of brightness in the J band. Data points of the source V1118 Ori are connected with a straight line.
   \label{LinevsJmag:fig}}
\end{figure}

\begin{figure}
 \centering
   \includegraphics [width=15 cm] {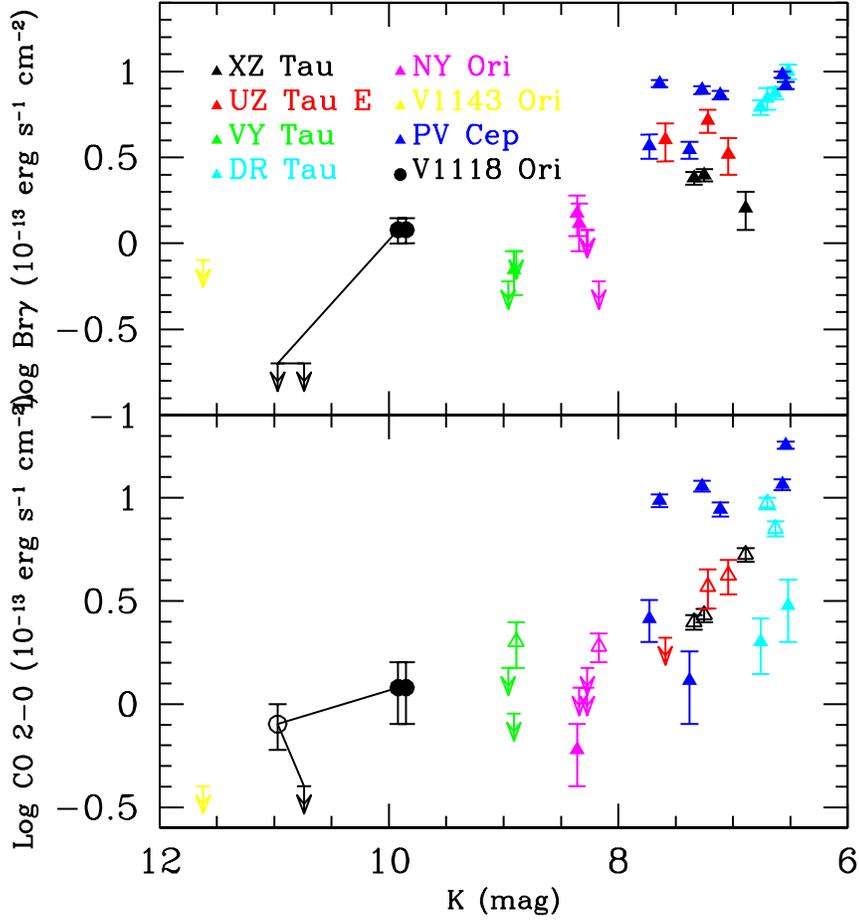}
   \caption{Observed Br$\gamma$ and CO 2-0 fluxes as a function of brightness in the K
   band. Open symbols indicate that CO line is in absorption.
   Data points of the source V1118 Ori are connected with a straight line.
   \label{LinevsKmag:fig}}
\end{figure}

The initial results of our ongoing survey are plotted in
Figures~\ref{LinevsJmag:fig} and \ref{LinevsKmag:fig}, where the
fluxes of few prominent lines observed in different dates are
depicted as a function of the source brightness in the
corresponding band (J for Pa$\beta$ and Pa$\gamma$ and K for
Br$\gamma$ and CO (2-0)). We remind that spectroscopy and
photometry are contemporary. Together with the new results, we
include for completeness also our data of V1118 Ori as anticipated
above. By examining both Figures, we see that the H recombination
lines plots show a similar behaviour, while the CO 2-0 correlation
with the K mag appears quite different since this feature is
detected both in emission (solid symbols) and in absorption (open
symbols) for different continuum levels . The recombination lines
exhibit an overall trend, according to which the brightest line
fluxes are associated to brightest sources. Moreover, some sources
(VY Tau, DR Tau, NY Ori and PV Cep) present line fluxes which tend
to increase as the continuum increases. This consideration seems
not to be applicable to all the individual sources; indeed, XZ Tau
and UZ Tau show constant line fluxes while the source brightens,
although by a small amount (less than 0.5 mag). Two interesting
cases are represented by V1118 Ori and PV Cep. Both sources
exhibit a large continuum variation ($\Delta$J \gapprox 1.2 mag,
$\Delta$K \gapprox 1.0 mag the former, and $\Delta$J \gapprox 2.2
mag, $\Delta$K \gapprox 1.2 mag the latter) associated with a
significant (more than a factor of 6 and 2, respectively) flux
variation of the recombination lines. Noticeably, V1118 Ori is the
only source deliberately monitored for a longer time during a
post-outburst phase, while monitoring the fading phase of PV Cep
has been fortuitous, since this source belongs to the targets of
our unbiased monitoring accomplished so far on a period of only
one year, in principle too short to document significant
variations. The sources for which the largest continuum variations
(more than one mag) have been sampled, are those presenting a
definite line flux increase as the continuum increases;
conversely, objects showing smaller continuum fluctuations (less
than half mag), show more erratic line variations and cannot be
directly correlated to continuum variations, but are more likely
related to local instabilities. Our preliminary conclusions need
to be confirmed by continuing our IR monitoring in order to
adequately sample larger variations (both in continua and lines)
whose existence, according to literature data (Paper II), is well
documented. The following preliminary analysis will be essentially
based on a restricted sub-sample of sources that present at the
moment a significant number of HI recombination lines, namely
those for which a reasonable fitting can be applied (see next
Sect. 4.2.1). This small sample is constituted by the spectra
taken in different epochs of UZ Tau, DR Tau, V1118 Ori, and PV
Cep.\\

\begin{figure}
 \centering
   \includegraphics [width=15 cm] {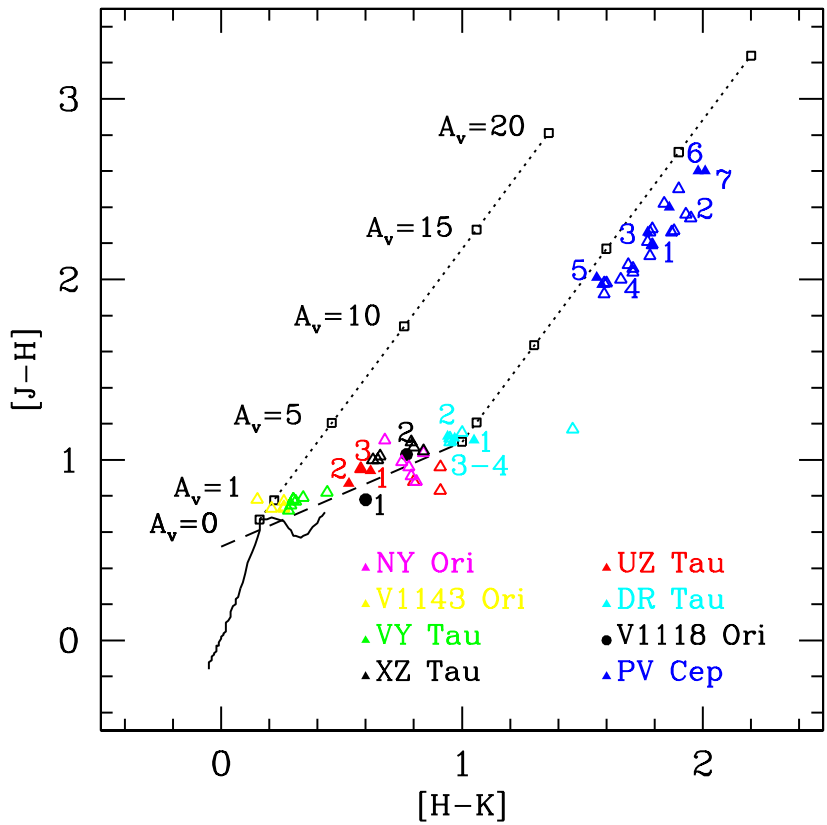}
   \caption{Near-IR two colours diagram of a selected sub-sample of EXor (see text) in different epochs.
   The solid line marks the unreddened main sequence, whereas the dashed one is the locus of the T Tauri stars
   (Meyer et al. 1997). Dotted lines represent the reddening law (Rieke \& Lebofsky 1985) where different intervals
   of A$_V$ (in mag) are indicated by open squares.
   \label{twocolours:fig}}
\end{figure}

Whatever is (are) the mechanism(s) responsible for line emission,
it is worthwhile to investigate whether extinction has a role in
determining the photometric and spectroscopic variability of
EXor's. To that scope, the JHK photometry contemporary to the
spectra (see Table~\ref{log:tab}) is provided in form of two
colours diagram. The colour variations of the four sources of our
sub-sample are depicted in Figure~\ref{twocolours:fig} and marked
with progressive numbers pertaining to different epochs. For
comparison purposes, in the same plot all the results of our
photometric monitoring are displayed: the solid (open) symbols
indicate that a contemporary near-IR spectrum exists (is absent).
Figure~\ref{twocolours:fig} gives also the extinction vector
(Rieke \& Lebofsky 1985) starting from both the unreddened main
sequence (solid line) and the locus of T Tauri stars (dashed line
- Meyer et al. 1997).

The objects that during our monitoring period have shown the
smallest fluctuations (ie. UZ Tau E and DR Tau) are, as expected,
associated to negligible color variations, apparently not related
to any extinction variation. Out the two sources showing the
largest photometric variations (V1118 Ori and PV Cep), the former
presents some indications in favour of an extinction dependance,
but we think this occurrence may be fortuitous since our prior
photometric monitoring evidenced an erratic change of position on
the same two colours plot (Fig.2 of Paper I). Moreover, if the
extinction variation of V1118 Ori between the two epochs (ie.
$\Delta$A$_V$ $\simeq$ 2.5) were genuine, it should not be enough
to account for the observed Br$\gamma$ fading by more than a
factor of six. The colours of the remaining EXor monitored so far
do not show any extinction dependance either.

The sole source for which a variable extinction seems to be the
most plausible reason for its photometric and spectroscopic
fluctuations is PV Cep. By comparing the photometric data
(Table~\ref{log:tab}) and its consequent location in
Figure~\ref{twocolours:fig} along the extinction vector, one can
see that all the bright phases correspond to diminishing A$_V$
values, while fading is always associated with an A$_V$
increasing. More quantitatively, A$_V$ has varied between 9.0 and
14.5 mag during our monitoring period (see also
Table~\ref{models:tab}). We have corrected the observed line
fluxes (just Pa$\beta$ and Br$\gamma$ for having a check) by using
the adequate A$_V$ as derived from the photometry of that epoch
(Figure~\ref{twocolours:fig}): by doing so, the intrinsic
Pa$\beta$ and Br$\gamma$ fluxes relative to different dates become
substantially equal (within a factor lesser than 1.5) and the
intrinsic line ratios Pa$\beta$/Br$\gamma$ remain all in the range
2.5-3.5 (Table~\ref{models:tab}). This result gives strong support
to the hypothesis that line flux variation in PV Cep are due to a
variable extinction more than to modification in the accretion (or
mass loss) rate.

Such A$_V$ variation can be translated into the volume density
needed to pass from A$_V$ = 9.0 to A$_V$ = 14.5 through the
relationship giving the column density, N(H$_2$)= A$_V$
10$^{21}$/R cm$^{-2}$, where R represents the size of the
obscuring matter. Even assuming a size equal to 5 stellar radii of
a M0 star, we would obtain a volume density n $\simeq$ 4 10$^9$
cm$^{-3}$, a value surprisingly high if it were uniformly
distributed on such a large volume. Instead, a similar amount of
dust can be supposed as organized in much smaller structures, such
as disk inhomogeneities, that repeatedly cross the line of sight.
Indeed, the time elapsed from maximum to minimum light (80 days)
is largely enough to cover many tens of stellar radii by
travelling at the Earth velocity.

The fact that PV Cep is a peculiar young object was already stated
many years ago (Cohen et al. 1981), through studies of the varying
bi-conical nebula associated to the star (GM29 - Gyul'budagyan \&
Magakyan 1977). This object shows simultaneously continuum
variations due to extinction and episodic mass ejection phenomena.
Indeed, the fan-shaped morphology of the varying nebula and the
strongest P-Cygni profiles in the hydrogen lines at the maximum
light suggest, respectively, that a circumstellar disk
intermittently obscurates the star and that the greatest
brightness may occur close to the highest mass ejection events.
Cohen et al. (1981) presented also optical continuum variations of
PV Cep, whose time-scales and amplitudes are well in agreement
with our observations. Further support to the disk presence is
given by the ice absorption feature around 3 $\mu$m that has been
interpreted as a common manifestation associated to edge-on
morphologies (van Citters \& Smith 1989).

The peculiarity of PV Cep (with respect to the other EXor's) displayed by our data
and confirmed by the literature, suggests some caveat in dealing with it as a confirmed
member of the class. Maybe PV Cep could be an EXor in a less evolved stage as other
candidates seem to be (see Sect. 4.5).

\subsection{HI recombination}

HI recombination lines largely dominate our near-IR spectra of
EXor, hence we will rely on them to probe the gas emitting region
and to discuss the correlation between line emission and continuum
variability.

The debate whether near-IR hydrogen lines observed in T Tauri
stars originate in the winds or arise in the accretion regions is
still open: the former is supported by: (i) the univocal presence
of winds; (ii) the presence of P-Cygni profiles; (iii) recent
spectro-astrometric studies of T Tauri stars (Whelan, Ray \& Davis
2004): (iv) the same magnetospheric accretion models. Analogously,
also the mass accretion scenario (e.g. Hartmann, Hewett \& Calvet
1994) sits on theoretical (e.g. Shu et al. 1994) and observational
(e.g. Kenyon et al. 1994) basis. Winds and accretion flows may
concur to originate the hydrogen line emission: both environments
are opaque to the Lyman continuum photons, but hydrogen is
photoionized, in any case, by the Balmer continuum photons (Natta
et al. 1988; Basri \& Bertout 1989). Whatever is the emission
region (infall envelope or wind), it is worthwhile noting that the
powering mechanism is likely the same, namely the accretion which
onsets the wind.\\

\begin{figure}
 \centering
   \includegraphics [width=10 cm] {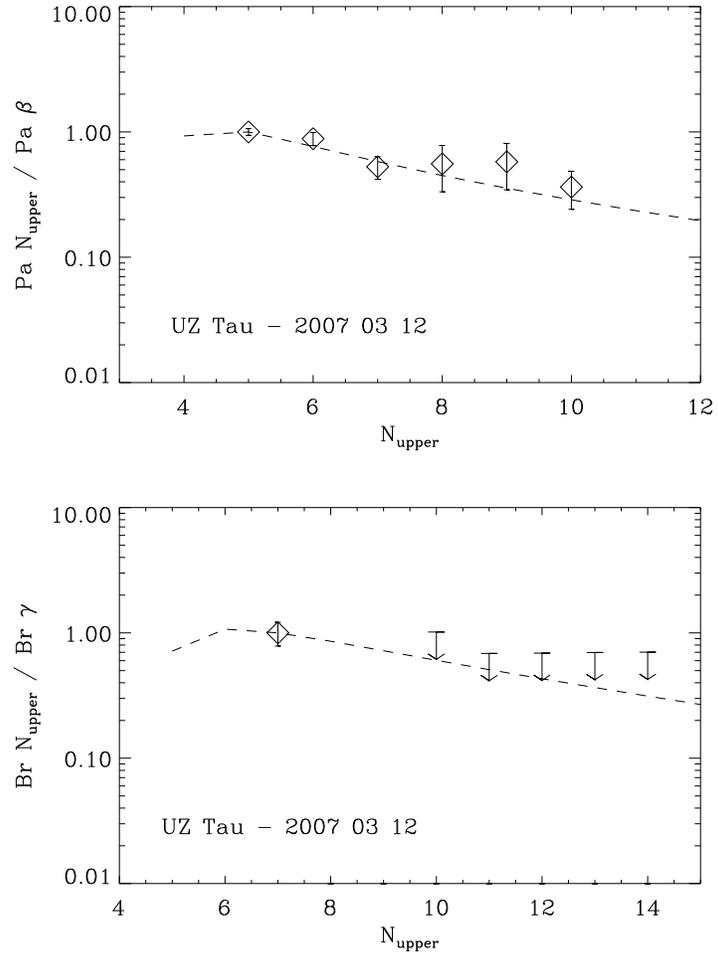}
   \caption{Line ratios of the Paschen (top) and Brackett (bottom) series for UZ Tau obtained
   in the indicated date. The best fit (dashed line) parameters are given in Table~\ref{models:tab}.
   \label{UZ:fig}}
\end{figure}

\begin{figure}
 \centering
   \includegraphics [width=10 cm] {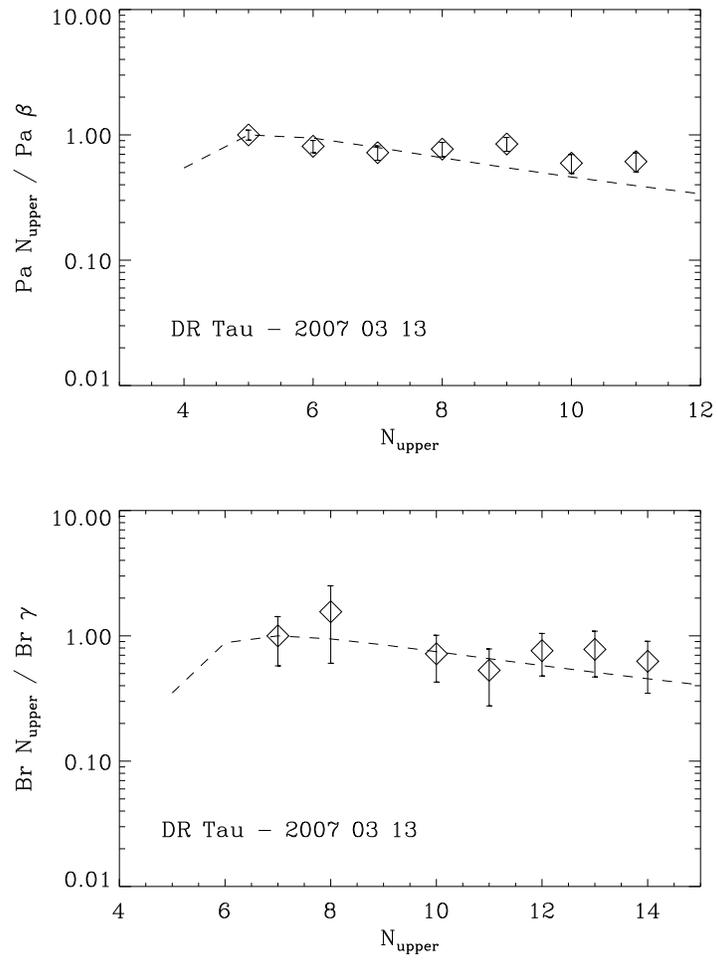}
   \caption{As Figure~\ref{UZ:fig} for DR Tau.
   \label{DR:fig}}
\end{figure}

\begin{figure}
 \centering
   \includegraphics [width=10 cm] {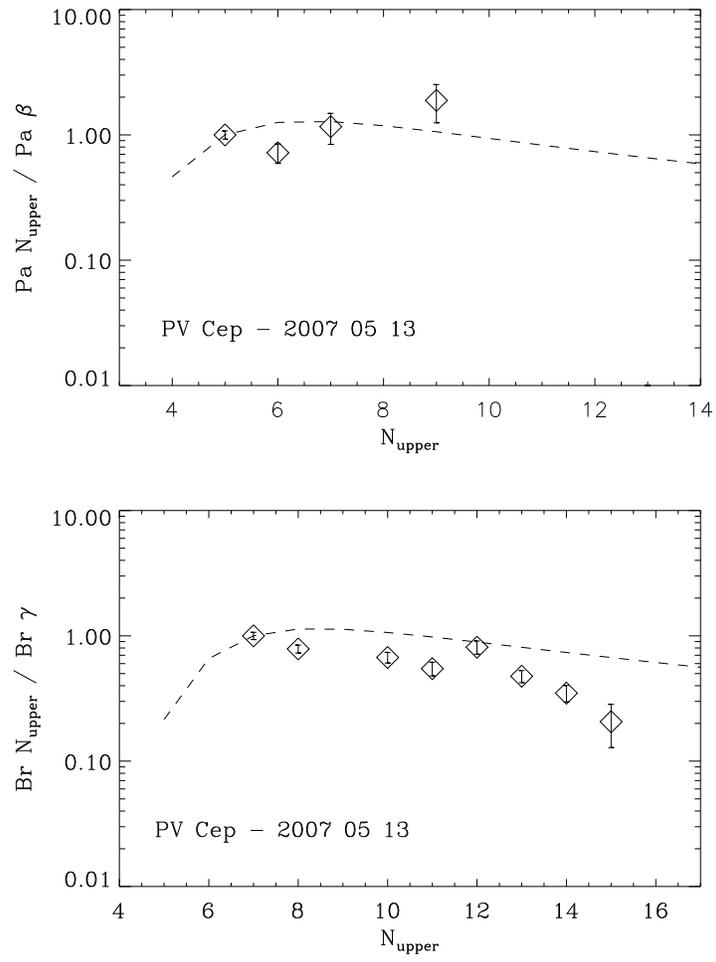}
   \caption{As Figure~\ref{UZ:fig} for PV Cep.
   \label{PV:fig}}
\end{figure}

In the framework of the wind hypothesis, we model the behaviour of
those sources where a significant number of HI recombination lines
have been observed with a wind model (Nisini, Antoniucci \&
Giannini 2004) that considers a spherically symmetric envelope
with a constant rate of mass loss ($\dot{M}$ = 4$\pi r^2 \rho(r)
v(r)$). The emitting gas is assumed to be in LTE and the adopted
gas velocity law is:

\begin{equation}
{v(r) = {v_i + (v}_{max} - v_i)[1 - (R_{*}/R)^{\alpha}]}
\end{equation}

\noindent where $v_i$ and $v_{max}$ are the initial and maximum
wind velocities, respectively, while $R_{*}$ is the stellar
radius. The best fit to the data points is obtained for the set of
parameters, namely the envelope size R$_{out}$ (expressed in units
of stellar radii $R_{*}$) abd the gas temperature T, given in
columns 4 and 5 of Table~\ref{models:tab}, respectively. We have
assumed input values equal for all the objects for the remaining
parameters, they are: the initial gas velocity $v_i$ = 30 km
s$^{-1}$; the envelope's internal radius $R_{i}$= 1 $R_{*}$, where
$R_{*}$ has the value corresponding to the stellar spectral type;
and the exponent $\alpha$ = 0.2. The adopted A$_V$ values (col.2
of Table~\ref{models:tab}) are taken from the literature
(Table~\ref{parameters:tab}) or from our photometry. For the
investigated sources, some model fitting to the data are depicted
in Figures~\ref{UZ:fig} to \ref{PV:fig}, as an example;
observational data are shown as line ratios of the Paschen and
Brackett series with respect to the Pa$\beta$ and Br$\gamma$,
respectively.

To evaluate the uncertainties on the derived parameters and to
check the sensitivity of our model, we computed the range of
variation for each input parameter that is allowed to eventually
provide line flux predictions comparable (within a 50 \% extent)
to the observed values. Such analysis indicates that gas
temperature is one of the less sensitive parameters: variations
between 5000 K and 10000 K do not affect the fit significantly.
Conversely, other parameters are quite critical and their
variability ranges are consequently rather narrow: about 50 \% for
$\dot{M}$, R$_i$ = 1 - 2 R$_{*}$, 50 \% for the envelope
thickness. Table~\ref{models:tab} gives, for each date, the ratio
between the observed (corrected for extinction) and predicted (by
our model) values of Pa$\beta$ (col.6) and Br$\gamma$ (col.7).
Pa$\beta$/Br$\gamma$ ratios, both predicted (col.8) and observed
(col.9) are listed, as well. Such a comparison (observations vs.
model) allows us to verify the goodness of the obtained fit. Due
to the uncertainties of the observations and even more to those of
modelling, we assume that fit to the data can be considered as
acceptable when ({\it i}) the ratios Pa$\beta$(mod/obs) and
Br$\gamma$(mod/obs) range between 0.7 and 1.5; and ({\it ii}) the
ratio Pa$\beta$/Br$\gamma$ derived from the model does not differ
from the observed one by more than 50\%. Such requirements are
well fulfilled in all cases listed in Table~\ref{models:tab}. The
sole exception is represented by some values of PV Cep (see the
caveat in Sect.4.1), which indeed shows observed ratios
Pa$\beta$/Br$\gamma$ lower than those of other sources and not
compatible with any reasonable combination of input parameters.
Finally column 10 of Table~\ref{models:tab} lists the mass loss
rate ($\dot{M}_{wind}$) derived from our model: these values range
between 10$^{-8}$ and 10$^{-7}$ M$_{\sun}$ yr$^{-1}$ (apart two
determinations for PV Cep), namely they are consistent with the
range of values inferred by independent determinations on active T
Tauri stars (Hartigan, Edwards \& Ghandour 1995).

Pa$\beta$ and Br$\gamma$ line luminosity can be also independently
used to get an estimate of the $\dot{M}_{acc}$ to be compared with
$\dot{M}_{wind}$ derived by our model. To that scope, the
empirical relationships given by Muzerolle, Hartmann \& Calvet
(1998) can be used. It relates the emission line (Pa$\beta$ and
Br$\gamma$) luminosities with the accretion luminosity as measured
from the continuum excess. We remark how the empirical nature of
such relationships allows us their applications whatever is the
origin of the recombination lines (accretion or wind).
Table~\ref{modelsacc:tab}, for each date, provides: the positive
or negative luminosity variation (integrated just over the JHK
bands - col.3) with respect to luminosity value obtained in the
first epoch; the extinction correction applied for deriving the
intrinsic fluxes (col.4); the accretion luminosities values
derived from the equations (1) and (2) of Muzerolle et al. (1998)
(col.5 and 6); From the accretion luminosity an estimate of the
mass accretion rate can be derived (col.7) (Gullbring et al. 1998)
by assuming the inner radius of the accretion disk R$_i$ (R$_i$ =
5 R$_{*}$), the stellar mass (M$_{*}$), and the stellar radius
(R$_{*}$) according to the spectral type indicated for each source
in Table~\ref{parameters:tab}. The derived mass accretion rates
roughly exceed by an order of magnitude the mass
loss ones, as predicted by wind models (see Sect.1).\\

Therefore we can conclude that the wind scenario alone is fully
consistent with the observational data. However, we are not able
to rule out that the same HI lines can be originated by accretion
alone. Indeed, this latter can account for the observed
Pa$\beta$/Br$\gamma$ ratios (roughly between 3 and 4, see
Table~\ref{models:tab}) only for $\dot{M}_{acc}$ $>$ 10$^{-8}$
M$_{\sun}$ yr$^{-1}$ (Muzerolle, Calvet \& Hartmann 2001 - their
Figure 15). Unfortunately, accretion models able to quantitatively
predict in a consistent way also wind emission do not exist,
hence, within this scheme, it is not easy to verify any
consistence between $\dot{M}_{wind}$ and $\dot{M}_{acc}$.

\subsection{CO overtone emission}

CO emission is seen together with Br$\gamma$ line emission, but
these two features probably come from different volumes of gas. At
temperatures of about 4000 K, both CO and molecular hydrogen are
dissociated by collisions. However, for density values higher than
10$^7$ cm$^{-3}$ and in the presence of H$_2$, CO dominates the
cooling (Scoville et al. 1980). Therefore CO bands are specific
probes of the circumstellar portions where the gas is relatively
warm at high densities. Recently interferometric observations of
CO emission in PMS objects have confirmed that CO comes from a
small region ($<$ 1 AU), i.e. the inner gaseous disk (Tatulli et
al. 2008), much more compact than that typical of the Br$\gamma$
emission, i.e. the stellar wind (Malbet et al. 2007). However,
young stars emitting Br$\gamma$ from regions more compact than the
dust sublimation radius, have been found, as well (Kraus et al.
2008).

A number of models have been proposed for producing CO emission
(for a summary see: Biscaya et al. 1997). Carr (1989) investigated
an accretion disk and a neutral stellar wind scenario. Inner
circumstellar disk regions (Najita et al. 1996), or infalling
material heated by adiabatic compression from 3000 to 6000 K
(Martin 1997), have been also proposed as regions where CO
emission may arise.

However, as anticipated in Sect.3, CO overtone features [typically
(2-0) and (3-1)] behave in a quite variable manner in EXor. With
reference to Figure~\ref{LinevsKmag:fig} (bottom panel) the
spectra of some sources (DR Tau, V1118 Ori and NY Ori) present CO
bands in emission (solid symbols) or in absorption (open symbols)
on different epochs. For the remaining sources, CO bands, although
largely variable, have been detected always in emission or always
in absorption, during our monitoring period. Trying to understand
whether CO features are related to the continuum emission, we
concentrate our attention on the couple of sources (V1118 Ori and
PV Cep) that exhibit the largest (i.e the most significant)
continuum variations ($\Delta$K $>$ 1 mag). While brightening,
V1118 Ori presents CO features passing from absorption to emission
and PV Cep presents emission features of increasing intensity.
This trend is the same described above (see Sect.1) for V1647 Ori,
the eruptive variable recently discovered.

A plausible interpretation of our observations can be given in
terms of two prevailing mechanisms. If we assume that CO
absorption is originated in the stellar photosphere (ref.), then
it can be detected only during the more quiescent phases when the
accretion rate is low, the continuum emission is low as well, and
the star surface is more easily visible. When the accretion rate
increases, it produces a significant increase of the UV radiation
at the accretion shock that in turn heats the inner disk favoring
CO emission associated to an enhanced continuum emission. It is
worthwhile to continue our spectroscopical monitoring to confirm
or not the proposed scenario. A further evidence that stellar
surface is more easily visible at minimum brightness was already
presented in Fig.6 of Paper II where we noticed how, near the
minimum, higher and more variable values of polarization were
detected, suggesting that, in such conditions, we can see the
heavy spotted and magnetized photosphere.

\subsection{NaI feature at 2.206 $\mu$m}

Given its low first ionization potential (5.1 eV), sodium can be
present close to low mass late type stars (like EXor, T Tauri) and
in regions capable to shield direct ionizing photons from earlier
type stars (e.g. circumstellar disks). The ionization structure of
the emitting region can be suitably traced by the flux ratio
Br$\gamma$/NaI, which is free by extinction effects because of the
closeness in wavelength between the two permitted lines.
Hydrogen is expected to be neutral in Na$^+$ region, thus a low value
of the Br$\gamma$/NaI ratio suggests a lower ionization; it might
indicate the presence of high density regions where photons cannot
easily penetrate.\\

\begin{figure}
 \centering
   \includegraphics [width=15 cm] {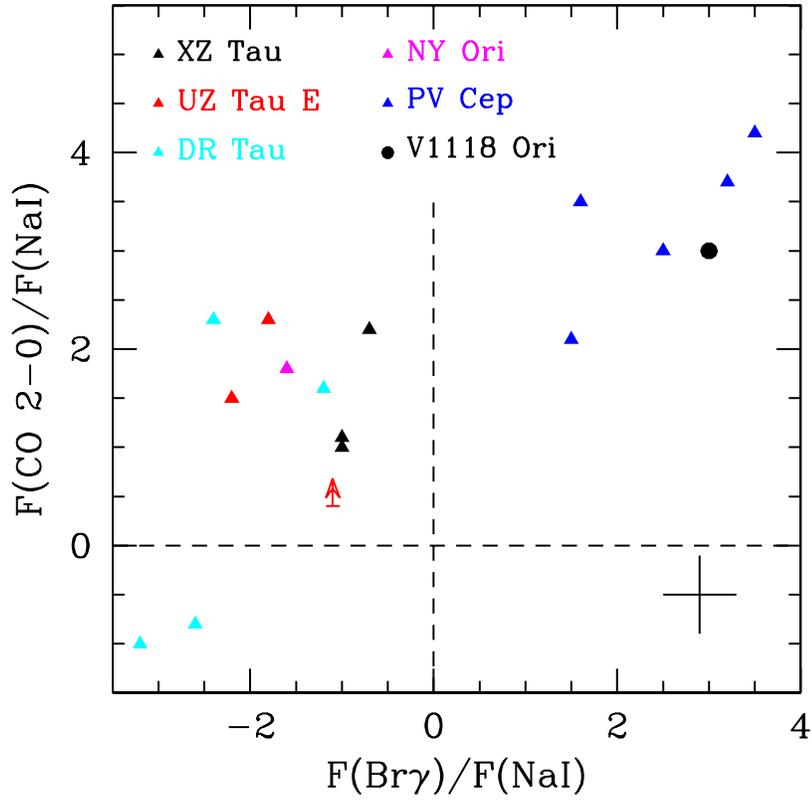}
   \caption{Observed ratios CO 2-0/NaI vs Br$\gamma$/NaI. Dashed lines delimit the loci where the line ratios are
   positive or negative. The cross on the bottom right corner represents the typical error, obtained
   by propagating the average uncertainty of the individual line fluxes.
   \label{CO_Na_ratios:fig}}
\end{figure}

In the presented EXor spectra the NaI 2.206 $\mu$m unresolved
doublet is detected both in emission and in absorption and,
remarkably, this line follows the same behaviour than that of the
CO bands. This circumstance is depicted in
Figure~\ref{CO_Na_ratios:fig}, where the ratio CO 2-0/NaI is
plotted vs the Br$\gamma$/NaI ratio for the different spectra
taken in different monitoring phases. Because of the closeness of
the three transitions (2.166, 2.206 and 2.293 $\mu$m Br$\gamma$,
NaI and CO 2-0, respectively), also this plot is unaffected by
extinction. The CO 2-0/NaI values are all positive (two spectra of
DR Tau represent the only exception), this means that NaI and CO
are seen either both in emission or both in absorption. Since
Br$\gamma$ is detected always in emission, the NaI emission cases
are located in the first quadrant (upper right) while NaI
absorption are those in the second quadrant (upper left). Such a
behaviour speaks in favour of a common origin for both NaI and CO
features. Hence, following the interpretation given for CO
features (Sect. 4.2), NaI line as well can be originated at the
star photosphere (absorption) or in the inner disk (emission),
depending on the outbursting stage of the object. Confirmation
that CO (2-0) and likely fluorescent NaI emission at 2.206 $\mu$m
can arise in the same low excitation region is provided by
McGregor et al. (1988), who note that carbon monoxide and sodium
require similar temperatures to survive: the former is
collisionally excited and emits at 3000-4000 K, as discussed in
Sect.4.2; the latter is shielded in the same high density
environment and is excited by adequate pumping photons without
radiatively ionizing NaI.

Figure~\ref{CO_Na_ratios:fig} shows how the values of the ratio
F(CO 2-0)/F(NaI) corresponding to emission cases are
systematically larger (they cluster in the range 2-4) than those
of absorption cases (between 1 and 2). This trend is likely
related to the different physical conditions between the stellar
photosphere and the circumstellar environment.

Antoniucci et al. (2007) noticed that the ratio Br$\gamma$/NaI for
jet driving sources ranges between 2 and 5, while larger values
are associated to Class I sources that show no evidence of jet
emission. The ratios we have obtained for the EXor (see
Figure~\ref{CO_Na_ratios:fig}) are low and in the same range of
the former group, although the presence of a jet does not seem a
property applicable to the EXor class. A plausible explanation is
that the low value of ratio Br$\gamma$/NaI is typical of those
cases where the role of the circumstellar disc dominates. NaI IR
emission lines, be they associated to jet driving sources or to
EXor variables, come from large columns of warm and neutral
material located in the inner part of the disc.

Finally, we note that all the objects that present NaI in
absorption have also CaI in absorption at 2.267 $\mu$m, at a
comparable intensity level. This provides further support to the
hypothesis that the stellar photosphere appears during the most
inactive phases.

\subsection{Comparison with spectra of sources candidate to be EXor}

In the recent years seven additional outbursting sources have been
tentatively recognized as EXor, although the attribution to this
class (or to the FUor one) is still debated. They are listed in
Table~\ref{candidates:tab} along with some relevant parameters.
The available near-IR spectra, apart one
featureless exception, are all emission line spectra strictly
resembling those of the EXors presented in Sect.3: in fact they
are largely dominated by hydrogen recombination along with other
ionic contributions and CO bandhead emission around 2.3 $\mu$m.
Moving from active to inactive phases line emission tends to fade
and the P$_{Cyg}$ profiles progressively diminish. Other feature
in common with the classical EXor is the photometric behaviour
generally becoming bluer while brightening
(Paper II). All in all candidates and confirmed members appear
quite equivalent with respect to their spectroscopic properties; a
substantial difference is that the candidates are more embedded
(A$_V$ \gapprox 10 mag) than the other EXor, suggesting, not
surprisingly, that this evolutionary stage might also occur earlier
than the T Tauri phase.

\section{Concluding Remarks}

We have presented the first results of a long term spectroscopic monitoring of EXor
in the near-IR aimed to investigate the variability in the accretion process, hence
modalities and timescales of the process itself. By analyzing the results, the following
can be summarized:

\begin{itemize}
\item[-] The presented part of our database refers just to the
starting period (2007-2008): more than 25 spectra have been obtained in different epochs
and they correspond to 8 sources in total. To correlate
continuum and line variability, all the observations are taken by performing
simultaneous photometry (JHK bands) and low resolution ($\mathcal{R}$ $\sim$ 250)
spectroscopy in the near-IR (0.8 to 2.5 $\mu$m).
\item[-] EXors near-IR spectra are line emission spectra dominated
by hydrogen recombination (Paschen and Brackett series) signalling
the presence of an ionized region. CO overtone features v=2-0,
v=3-1 are commonly detected both in emission and in absorption and
weaker atomic features (NaI and CaI) appear in emission or
absorption following the CO behaviour. At our level of
sensitivity, molecular hydrogen contributions are absent,
suggesting that shocks do not represent a major mechanism of
excitation. All in all EXor IR spectra look like those of the
accreting T Tauri stars more than those of FUor objects.
\item[-] Dealing with the first results of a long term monitoring
program, an attempt has been done to complement our data with
literature spectra so far collected. The last 30 years of near-IR
spectroscopic results of EXor, prior to our survey, consist in
just 14 spectra. These very few and sparse observations only
suggest that some line emission variability exists on very
uncertain timescales. Complementing these data with the results of
our survey, we can estimate an IR line variability from a factor
of 2 up to an order of magnitude and typical timescales from
months to years, respectively.
\item[-] The sources for which the largest continuum variations (more than one mag) have been sampled,
are those presenting a well defined HI line flux increase as the continuum increases;
conversely, objects showing smaller continuum fluctuations
(less than half mag), show more erratic line variations and cannot be straightforwardly associated to well
characterized accretion events, but are more likely related to local instabilities.
\item[-] The possibility that line variability is due to a variable extinction has been explored, and appears
well applicable only to the spectra taken in different epochs of the peculiar source PV Cep.
\item[-] The obtained spectra have been compared with a wind model
that considers a spherically symmetric and partially ionized
envelope with a constant rate of mass loss. Mass loss rates in the
range 2-10 10$^{-8}$ M$_{\sun}$ yr$^{-1}$ are derived along with
additional wind parameters. A possible origin of HI lines from
accretion is investigated and it also provides consistent results.
\item[-] CO overtone features have been detected both in emission and in absorption.
They are highly variable even on a short (days to months) timescale. In particular,
while brightening, one source presents CO features passing from absorption to
emission and another presents emission features of increasing intensity. CO
absorption could be originated in the stellar photosphere, thus
it can prevail only during quiescent phases when the star surface is more easily visible.
When the accretion rate increases the UV radiation at the accretion shock heats the
inner disk favoring CO emission.
\item[-] NaI 2.206 $\mu$m (and more rarely CaI 2.267 $\mu$m) is detected both in
emission and in absorption and, remarkably, it follows what CO bands do. Consequently
all these lines are thought to arise in the same regions: the photosphere (when in absorption),
or the inner disk (when in emission).
\item[-] Finally, a comparison is made with the outbursting sources
tentatively recognized as candidate EXor. They appear quite equivalent
with respect to their spectroscopic properties; a
substantial difference is that the candidates are more embedded
than the other EXor, suggesting that this evolutionary stage might also occur earlier
than the T Tauri phase.
\end{itemize}

A systematic investigation of the quantitative
(inter-)relationship between line and continuum flux variations
will be possible once our survey will achieve a longer coverage
and hence a larger number of significant accretion events could be
sampled.

\begin{deluxetable}{lccccccccccc}
\tabletypesize{\footnotesize} \tablecaption{EXOrs observed
parameters.\label{parameters:tab}} \tablewidth{0pt} \tablehead{
\colhead{Target} & \colhead{Dist.} &  \colhead{A$_V$} &
\colhead{L$_{bol}$}
    &\colhead{SpT} & \colhead{V$_{max}$} & \colhead{V$_{min}$} &
    \colhead{outf.} &  \colhead{location} & \colhead{binary} & \colhead{Ref.}
} \startdata
       & (pc)  & (mag)  & ($L_{\sun}$) &     & \multicolumn{2}{c}{(mag)} &  &   &  &    \\
\tableline
XZ Tau      &  140 &  3: &4.2 - 10.7& M3 & 10.4 & 16.6   &  Y  & L1551  &  Y (0.30)  &1,2,3,4,5,6 \\
UZ Tau E    &  140 & 1.49&   1.7    & M1,3& 11.7  & 15.0 &     & B19    &  Y (0.34)  & 1,3        \\
VY Tau      &  140 & 0.85&   0.75   & M0V & 9.0   & 15.3 &     & L1536  &  Y (0.66)  & 1,3,7      \\
DR Tau      &  140 &1.7-2.1&1.05-5.0&K7-M0& 10.5  & 16.0 &     & L1558  &  N         & 1,3,8,9    \\
V1118 Ori   &  414 & 0-2   &1.4-25.4& M1e & 14.2  & 17.5 &  N  & ONC    &  Y (0.18)  & 1,10,11,12 \\
NY Ori      &  414 & 0.3 &            & ??& 14.5& 17.5 &       & ONC    &  N         & 1,13,14    \\
V1143 Ori   &  460 &     &            & M   &  13 &  19  &     & L1640  &  -         & 1          \\
PV Cep      &  500 &0.4-5&   100      & A5e & 11.1& 18.0 &  Y  & L1158  &  -         & 1,15,16    \\
\enddata

\tablecomments{References to the Table: (1) SIMBAD Astronomical
Database (http://simbad.u-strasbg.fr/simbad); (2) Coffey, Downes
\& Ray 2004; (3) Herbig \& Bell 1988; (4) Carr 1990; (5) Evans II
et al. 1987; (6) Beck et al. 2008; (7) Herbig 1990; (8) Cohen \&
Kuhi 1979; (9) Kenyon et al. 1994; (10) Lorenzetti et al. (2006) -
Paper I; (11) Reipurth et al. 2007; (12) Menten et al. 2007; (13)
Breger, Gherz \& Hackwell 1981; (14) K\"{o}hler et al. 2006; (15)
Cohen et al. 1981; (16) van Citters \& Smith 1989.}
\end{deluxetable}

\begin{deluxetable}{lcccccl}
\tabletypesize{\footnotesize} \tablecaption{Log of our
observations: date on which spectra have been taken, near-IR
magnitudes at that epoch and references to literature near-IR
spectroscopic data.\label{log:tab}} \tablewidth{0pt} \tablehead{
\colhead{Target} & \colhead{Date} & \colhead{J} & \colhead{H} &
\colhead{K} & \colhead{Status$^a$} & \colhead{Ref. to near-IR
spectroscopy} } \startdata
          &           &\multicolumn{3}{c}{(mag)}& &        \\
\tableline
XZ Tau    &2007 Mar 08& 9.21& 8.14& 7.34& H &  near- (1); Pa$\beta$, Br$\gamma$ (2); Br$\gamma$,Br$\alpha$ (3); \\
          &2007 Oct 08& 8.93& 7.91& 7.25& Hr&  2-4$\mu$m (4); Br$\gamma$, H$_2$ (5); Pa$\beta$, Br$\gamma$ (6); \\
          &2008 Mar 28& 8.78& 7.68& 6.89& Hr&  CO bands abs.(7); HeI, Pa$\gamma$ (8);                           \\
          &           &     &     &     &   &  extended H$_2$ emission (9).                                                                 \\
\tableline
UZ Tau E  &2007 Mar 12& 8.57& 7.62& 7.04& H &  near- (10); HeI, Pa$\gamma$ (8)                                   \\
          &2007 Oct 11& 8.62& 7.75& 7.22& Hd&                                                                   \\
          &2008 Mar 30& 9.15& 8.21& 7.59& Hd&                                                                   \\
\tableline
VY Tau    &2007 Mar 14&10.00& 9.25& 8.96& H &  Pa$\beta$ (2); near- (1),(11); HeI, Pa$\gamma$ (8)               \\
          &2007 Oct 11&10.02& 9.23& 8.89& H &                                                                   \\
          &2008 Mar 30& 9.99& 9.21& 8.91& H &                                                                   \\
\tableline
DR Tau    &2007 Mar 13& 8.68& 7.57& 6.52& I &  near-(1); Pa$\beta$, Br$\gamma$ (2); HeI (12);                   \\
          &2007 Oct 08& 8.83& 7.70& 6.76& I &  Pa$\beta$, Br$\gamma$ (6),(13); H$_2$ rot.                       \\
          &2007 Nov 03& 8.72& 7.60& 6.63& I &                                                                   \\
          &2008 Mar 28& 8.76& 7.66& 6.70& I &                                                                   \\
\tableline
V1118 Ori &2005 Sep 11&11.23&10.45& 9.85& H &  near-, CO bands em.(14)/abs (17)                                 \\
          &2006 Sep 23&12.54&11.51&10.74& L &                                                                   \\
\tableline
NY Ori    &2007 Mar 15&10.10& 9.14& 8.36&   &                                                                   \\
          &2007 Oct 14&10.01& 9.02& 8.27&   &                                                                   \\
          &2007 Oct 15& 9.97& 9.06& 8.27&   &                                                                   \\
          &2007 Oct 24&10.22& 9.18& 8.34&   &                                                                   \\
          &2008 Mar 29& 9.96& 8.85& 8.17&   &                                                                   \\
\tableline
V1143 Ori &2007 Mar 17&  -  &  -  &  -  &   &                                                                   \\
          &2007 Oct 15&12.65&11.88&11.62& I &                                                                   \\
\tableline
PV Cep    &2007 May 13&11.52& 9.33& 7.54& L &  Br$\gamma$, H$_2$ (5); CO (7); [FeII] (15);                      \\
          &2007 Oct 15&11.53& 9.13& 7.27& Lr&  2-4$\mu$m (4)(16)                                                \\
          &2007 Oct 27&11.14& 8.88& 7.11& Lr&                                                                   \\
          &2008 Apr 01&10.09& 8.12& 6.54& Hr&                                                                   \\
          &2008 Apr 12&10.14& 8.13& 6.57& H &                                                                   \\
          &2008 Jun 10&11.96& 9.36& 7.38& Ld&                                                                   \\
          &2008 Jun 18&12.34& 9.74& 7.73& Ld&                                                                   \\
\enddata
\tablecomments{References to the Table: (1) Folha \& Emerson 1999;
(2) Folha \& Emerson 2001; (3) Evans et al. 1987; (4) Sato et al.
1990; (5) Carr 1990; (6) Giovanardi et al. 1991; (7) Biscaya et
al. 1997; (8) Edwards et al. 2006; (9) Beck et al. 2008; (10)
Elias 1978; (11) Greene \& Lada 1996; (12) Edwards et al. 2003;
(13) Muzerolle, Hartmann \& Calvet 1998; (14) Paper I; (15) Hamann
et al. 1994; (16) van Citters \& Smith 1989; (17) Paper II. }
\tablenotetext{a}{~~~~ H = high, I = intermediate, L = low, r =
rising, d = declining }
\end{deluxetable}

\begin{deluxetable}{cccccccc}
\tabletypesize{\footnotesize} \tablecaption{Line emission fluxes
of XZ Tau. \label{linesxztau:tab}} \tablewidth{0pt} \tablehead{
\colhead{$\lambda_{vac}$} & \colhead{Ident.} & \colhead{F $\pm$
$\Delta$F} & \colhead{EW} & \colhead{F $\pm$ $\Delta$F} &
\colhead{EW} & \colhead{F $\pm$ $\Delta$F} & \colhead{EW} }
\startdata
($\mu$m) &   & \multicolumn{2}{c}{07/Mar/08} & \multicolumn{2}{c}{07/Oct/08} & \multicolumn{2}{c}{08/Mar/28}\\
\tableline
0.923    &  Pa9        &  $<$ 11         &  -   &  $<$ 6        & -   &      -        & -   \\
0.955    &  Pa8        &      -          &  -   &  3.5 $\pm$ 1  & -5  & 2.1 $\pm$ 0.4 & -3  \\
1.005    &  Pa$\delta$ &  4 $\pm$ 1      &  -7  &  5 $\pm$ 1    & -7  & 1.8 $\pm$ 0.4 & -2  \\
1.094    &  Pa$\gamma$ &  4 $\pm$ 2$^*$  &  -8  &  3.5 $\pm$ 1  & -4  & 8 $\pm$ 4$^*$ & -8  \\
1.117    &  ?          &      -          &  -   &  4 $\pm$ 1    & -5  &      -        & -   \\
1.288    &  Pa$\beta$  &  6   $\pm$ 1    & -10  &  6 $\pm$ 2    & -7  & 6   $\pm$ 1   & -7  \\
1.328    &  ?          &  3.2 $\pm$ 0.6  & -5   &3.4 $\pm$ 0.6  & -4  & 1.8 $\pm$ 0.6 & -2  \\
2.166    &  Br$\gamma$ &  2.4 $\pm$ 0.2  & -4   &2.5 $\pm$ 0.2  & -4  & 1.6 $\pm$ 0.4 & -2  \\
2.208    &  NaI        & -2.5 $\pm$ 0.6  & +4   &-2.5 $\pm$ 0.6 & +4  &-2.4 $\pm$ 0.6 & +4  \\
2.267    &  CaI        & -2.6 $\pm$ 0.6  & +5   &-1.2 $\pm$0.6$^*$& +2&     $<$1.8    & -   \\
2.293    &  CO 2-0     & -2.5 $\pm$ 0.6  & +5   &-2.7 $\pm$ 0.6 & +5  &-5.3 $\pm$ 0.4 & +8  \\
2.323    &  CO 3-1     &     -           &  -   &    -          & -   &-4.4 $\pm$ 0.4 & +7  \\
\enddata

\tablecomments{Fluxes marked with an asterisk are those derived at
a 2$<$S/N$<$3 level; they will not be used in the following
analysis.\\
\noindent - Fluxes and errors are given in units of 10$^{-13}$erg
s$^{-1}$cm$^{-2}$, while EW are expressed in (\AA)\\
\noindent - The values of $\lambda_{oss}$ are not given since
their difference with the $\lambda_{vac}$ ones are always lesser
than the spectral resolution element.}

\end{deluxetable}

\begin{deluxetable}{cccccccc}
\tabletypesize{\footnotesize} \tablecaption{Line emission fluxes
of UZ Tau E. \label{linesuztau:tab}} \tablewidth{0pt} \tablehead{
\colhead{$\lambda_{vac}$} & \colhead{Ident.} & \colhead{F $\pm$
$\Delta$F} & \colhead{EW} & \colhead{F $\pm$ $\Delta$F} &
\colhead{EW} & \colhead{F $\pm$ $\Delta$F} & \colhead{EW} }
\startdata
($\mu$m) &   & \multicolumn{2}{c}{07/Mar/12} & \multicolumn{2}{c}{07/Oct/11} & \multicolumn{2}{c}{08/Mar/30} \\
\tableline
0.855    &  Pa15       &  9 $\pm$ 3       & -19  &  -                & -   &      -           & -    \\
0.867    &  Pa13       &  6 $\pm$ 2       & -12  &  -                & -   &      -           & -    \\
0.902    &  Pa10       &  3 $\pm$ 1       & -7   &  -                & -   &      -           & -    \\
0.923    &  Pa9        &  6 $\pm$ 1       & -16  &  -                & -   &      -           & -    \\
0.955    &  Pa8        &  6 $\pm$ 1       & -10  & 6 $\pm$ 1         & -5  & 15 $\pm$ 1       & -10  \\
1.005    &  Pa$\delta$ &  5 $\pm$ 1       & -7   & 6 $\pm$ 1         & -5  &      -           & -    \\
1.094    &  Pa$\gamma$ &  9 $\pm$ 1       & -11  & 9 $\pm$ 1         & -7  &  6 $\pm$ 1       & -4   \\
1.169    &   ?         &  4 $\pm$ 1       & -4   & 5 $\pm$ 1         & -5  &      -           & -    \\
1.282    &  Pa$\beta$  &  11.4 $\pm$ 0.8  & -13  & 10.4 $\pm$ 0.8    & -10 & 10 $\pm$ 1       & -8   \\
1.328    &   ?         &  5 $\pm$ 1       & -5   & 3 $\pm$ 1         & -3  &  3 $\pm$ 1       & -3   \\
1.491    &  Br27       &  4 $\pm$ 2$^*$   & -5   &  -                & -   &      -           & -    \\
1.588    &  Br14       &  4 $\pm$ 2$^*$   & -4   &  -                & -   &      -           & -    \\
1.611    &  Br13       &  4 $\pm$ 2$^*$   & -5   &  -                & -   &      -           & -    \\
1.641    &  Br12       &  4 $\pm$ 2$^*$   & -5   & 4 $\pm$ 2$^*$     & -5  &      -           & -    \\
1.681    &  Br11       &  4 $\pm$ 2$^*$   & -4   &  -                & -   &      -           & -    \\
1.694    &  ?          &  4 $\pm$ 2$^*$   & -4   &  -                & -   &      -           & -    \\
1.737    &  Br10       &  6 $\pm$ 3$^*$   & -7   &  -                & -   &      -           & -    \\
2.166    &  Br$\gamma$ &  3.3 $\pm$ 0.8   & -6   & 5.2 $\pm$ 0.8     & -9  &  4 $\pm$ 1       & -6   \\
2.208    &  NaI        &-1.8 $\pm$ 0.8$^*$& +3   &-2.4 $\pm$ 0.8     & +4  &  -3.7 $\pm$ 0.8  & +6   \\
2.267    &  CaI        & -2.4 $\pm$ 0.8   & +3   &-2.6 $\pm$ 0.8     & +5  &     $<2.4$       & -    \\
2.293    &  CO 2-0     & -4.2 $\pm$ 0.8   & +6   &-3.7 $\pm$ 0.8     & +7  &-1.5 $\pm$0.7$^*$ & +2   \\
2.323    &  CO 3-1     & -4.1 $\pm$ 0.8   & +6   &  -                & -   &      -           & -    \\
\enddata

\tablecomments{The same as in Table~\ref{linesxztau:tab}.}

\end{deluxetable}

\begin{deluxetable}{cccccccc}
\tabletypesize{\footnotesize} \tablecaption{Line emission fluxes
of VY Tau. \label{linesvytau:tab}} \tablewidth{0pt} \tablehead{
\colhead{$\lambda_{vac}$} & \colhead{Ident.} & \colhead{F $\pm$
$\Delta$F} & \colhead{EW} & \colhead{F $\pm$ $\Delta$F} &
\colhead{EW} & \colhead{F $\pm$ $\Delta$F} & \colhead{EW} }
\startdata
($\mu$m) &   & \multicolumn{2}{c}{07/Mar/14} & \multicolumn{2}{c}{07/Oct/11} & \multicolumn{2}{c}{08/Mar/30} \\
\tableline
1.094 & Pa$\gamma$  &  $<$ 1.8          & -  & 3 $\pm$ 1         & -11  & 3.1 $\pm$ 0.9     & -9   \\
1.282 & Pa$\beta$   & 1.9 $\pm$ 0.4     & -6 & 1.7 $\pm$ 0.4     & -6   & 1.9 $\pm$ 0.4     & -6   \\
1.327 &  ?          & 1.1 $\pm$ 0.5$^*$ & -4 & $<$1.2            & -    & 1.5 $\pm$ 0.4     & -6   \\
2.166 & Br$\gamma$  & 0.5 $\pm$ 0.2$^*$ & -7 & $<$ 0.9           & -    & 0.7 $\pm$ 0.2     & -6   \\
2.208 & NaI         &-1.2 $\pm$ 0.5$^*$ & +10& $<$ 1.8           & -    &-0.4 $\pm$ 0.2$^*$ & +2.5 \\
2.267 & CaI         & $<$ 1.5           & -  &    -              & -    &-0.5 $\pm$ 0.2$^*$ & +4   \\
2.293 & CO 2-0      &-0.8 $\pm$ 0.5$^*$ & +7 &-2.0 $\pm$ 0.5     & +19  &-0.6 $\pm$ 0.3$^*$ & +5   \\
\enddata

\tablecomments{The same as in Table~\ref{linesxztau:tab}. Upper
limits are given as 3$\sigma$ values.}

\end{deluxetable}

\begin{deluxetable}{cccccccccc}
\tabletypesize{\footnotesize} \tablecaption{Line emission fluxes
of DR Tau. \label{linesdrtau:tab}} \tablewidth{0pt} \tablehead{
\colhead{$\lambda_{vac}$} & \colhead{Ident.} & \colhead{F $\pm$
$\Delta$F} & \colhead{EW} & \colhead{F $\pm$ $\Delta$F} &
\colhead{EW} & \colhead{F $\pm$ $\Delta$F} & \colhead{EW} &
\colhead{F $\pm$ $\Delta$F} & \colhead{EW} } \startdata
($\mu$m) &   & \multicolumn{2}{c}{07/Mar/13} & \multicolumn{2}{c}{07/Oct/08} & \multicolumn{2}{c}{07/Nov/03} & \multicolumn{2}{c}{08/Mar/28}\\
\tableline
0.847    &  Pa17       &        -         &   -  &       -             &  -   &     13 $\pm$ 1  & -9     &          -          &  -     \\
0.855    &  Pa15       &    67 $\pm$ 0.8  & -39  &     50 $\pm$ 1      & -67  &     36 $\pm$ 1  & -37    &     75 $\pm$ 1      &  -73   \\
0.867    &  Pa13       &    29 $\pm$ 0.8  & -16  &     20 $\pm$ 1      & -27  &     34 $\pm$ 1  & -25    &     26 $\pm$ 1      &  -26   \\
0.886    &  Pa11       &    15 $\pm$ 0.8  & -7   &      5 $\pm$ 1      & -7   &     11 $\pm$ 1  & -9     &     16 $\pm$ 1      &  -16   \\
0.902    &  Pa10       &    15 $\pm$ 0.8  & -8   &      4 $\pm$ 2$^*$  & -5   &     15 $\pm$ 2  & -13    &     15 $\pm$ 1      &  -16   \\
0.923    &  Pa9        &    22 $\pm$ 0.8  & -11  &      9 $\pm$ 1      & -12  &     20 $\pm$ 1  & -16    &     16 $\pm$ 1      &  -17   \\
0.955    &  Pa8        &    21 $\pm$ 0.8  & -14  &      9 $\pm$ 1      & -11  &     21 $\pm$ 1  & -17    &     11.0 $\pm$ 0.5  &  -12   \\
1.005    &  Pa$\delta$ &    21 $\pm$ 0.8  & -13  &     12 $\pm$ 1      & -15  &     22 $\pm$ 1  & -18    &     13.4 $\pm$ 0.5  &  -13   \\
1.070    & ?           &    4.4 $\pm$ 0.8 & -4   &      3 $\pm$ 1      & -4   &      7 $\pm$ 1  & -6     &     3.7 $\pm$  0.5  &  -4    \\
1.094    &  Pa$\gamma$ &    26 $\pm$ 0.8  & -17  &     18 $\pm$ 2      & -22  &     25 $\pm$ 2  & -21    &    20.0 $\pm$  0.5  &  -20   \\
1.117    &  CI?        &    5 $\pm$ 1     & -4   &     5 $\pm$ 1       & -6   &       -         & -      &    2 $\pm$ 1$^*$    &  -2    \\
1.183    &  MgI        &    2 $\pm$ 1$^*$ & -1   &     2 $\pm$ 1$^*$   & -2   &    2 $\pm$ 1$^*$& -2     &    2 $\pm$ 1$^*$    &  -2    \\
1.282    &  Pa$\beta$  &    37 $\pm$ 0.8  & -24  &     24 $\pm$ 2      & -26  &     32 $\pm$ 2  & -29    &    30.0 $\pm$  0.5  &  -31   \\
1.328    &   ?         &    2.4 $\pm$ 0.4 & -2   &     -               & -    &       -         & -      &     3 $\pm$ 1       &  -4    \\
1.491    &  Br27       &   2.7 $\pm$ 0.9  & -4   &       -             &  -   &       -         &  -     &          -          &   -    \\
1.494    &  Br26       &   2.1 $\pm$ 0.2  & -4   &       -             &  -   &       -         &  -     &     2.2 $\pm$ 0.5   &  -2    \\
1.556    &  Br16       &         -        &  -   &       -             &  -   &       -         &  -     &     2.2 $\pm$ 0.5   &  -2    \\
1.570    &  Br15       &         -        &  -   &       -             &  -   &       -         &  -     &     5.9 $\pm$ 0.5   &  -4    \\
1.588    &  Br14       &   5.5 $\pm$ 0.6  & -6   &    3.1 $\pm$ 0.6    & -3   &    7.7 $\pm$ 0.6& -6     &     5.9 $\pm$ 0.5   &  -5    \\
1.611    &  Br13       &   6.9 $\pm$ 0.6  & -7   &    3.9 $\pm$ 0.6    & -4   &    4.7 $\pm$ 0.6& -4     &     3.4 $\pm$ 0.5   &  -3    \\
1.641    &  Br12       &   6.8 $\pm$ 0.6  & -7   &    2.7 $\pm$ 0.6    & -3   &    7.3 $\pm$ 0.6& -6     &     3.5 $\pm$ 0.5   &  -3    \\
1.681    &  Br11       &   4.8 $\pm$ 0.6  & -2   &    4.0 $\pm$ 0.6    & -3   &    7.2 $\pm$ 0.6& -6     &     8.5 $\pm$ 0.5   &  -7    \\
1.737    &  Br10       &   6.6 $\pm$ 0.6  & -7   &    4.5 $\pm$ 0.6    & -3   &    8.2 $\pm$ 0.6& -7     &     8.1 $\pm$ 0.5   &  -7    \\
1.745    &  HeI ?      &   7.1 $\pm$ 0.6  & -7   &    2.2 $\pm$ 0.6    & -2   &    3.4 $\pm$ 0.6& -3     &          -          &  -     \\
1.945    &  Br$\delta$ &   15 $\pm$ 5     & -13  &     16 $\pm$ 5      & -16  &    9.8 $\pm$ 0.5& -10    &    14.0 $\pm$ 0.5   &  -19   \\
2.059    &  HeI ?      &   2.9 $\pm$ 0.6  & -3   &    2.4 $\pm$ 0.6    & -2   &    2.1 $\pm$ 0.6& -2     &          -          &  -     \\
2.166    &  Br$\gamma$ &   10 $\pm$ 1     & -8   &    6.2 $\pm$ 0.6    & -6   &    7.5 $\pm$ 0.6& -7     &      7 $\pm$ 1      &  -6    \\
2.208    &  NaI        &   -3.1 $\pm$ 0.4 & +3   &   -2.4 $\pm$ 0.4    & +3   &   -3.1 $\pm$ 0.4& +3     &      -5.7$\pm$ 0.4  &  +5    \\
2.267    &  CaI        &   -2.5 $\pm$ 0.4 & +2   &   -1.0 $\pm$ 0.4$^*$& +1   &      -          & -      &              -      &  -     \\
2.293    &  CO 2-0     &    3 $\pm$ 1     & -3  &    2.0 $\pm$ 0.6     & -2   &   -7.1 $\pm$ 0.6& +7     &      -9.4$\pm$ 0.6  &  +9    \\
2.323    &  CO 3-1     &       -          &  -  &    2.4 $\pm$ 0.6     & -3   &   -7.2 $\pm$ 0.6& +7     &      -7.8$\pm$ 0.6  &  +7    \\
\enddata

\tablecomments{The same as in Table~\ref{linesxztau:tab}.}

\end{deluxetable}

\begin{deluxetable}{cccccccccccc}
\tabletypesize{\scriptsize} \tablecaption{Line emission fluxes of
NY Ori. \label{linesnyorise:tab}} \tablewidth{0pt} \tablehead{
\colhead{$\lambda_{vac}$} & \colhead{Ident.} & \colhead{F $\pm$
$\Delta$F} & \colhead{EW} & \colhead{F $\pm$ $\Delta$F} &
\colhead{EW} & \colhead{F $\pm$ $\Delta$F} & \colhead{EW} &
\colhead{F $\pm$ $\Delta$F} & \colhead{EW} & \colhead{F $\pm$
$\Delta$F} & \colhead{EW}} \startdata ($\mu$m) & &
\multicolumn{2}{c}{07/Mar/15} & \multicolumn{2}{c}{07/Oct/14} &
\multicolumn{2}{c}{07/Oct/15}
& \multicolumn{2}{c}{07/Oct/24} & \multicolumn{2}{c}{08/Mar/29}\\
\tableline
1.005 & Pa$\delta$ &       -       &  - & 1.9 $\pm$ 0.5     & -6 & 1.0 $\pm$ 0.5$^*$ & -3 &       -           &  - &     -         & -   \\
1.094 & Pa$\gamma$ & 2.5 $\pm$ 0.5 & -8 & 3.0 $\pm$ 0.5     & -9 & 1.5 $\pm$ 0.5     & -5 & 1.2 $\pm$ 0.5$^*$ & -4 & 1.7 $\pm$ 0.5 & -5  \\
1.282 & Pa$\beta$  & 3.2 $\pm$ 0.5 &-11 & 4.0 $\pm$ 0.5     & -13& 2.5 $\pm$ 0.5     & -8 & 2.2 $\pm$ 0.5     & -7 & 3.3 $\pm$ 0.5 & -10 \\
2.166 & Br$\gamma$ & 1.5 $\pm$ 0.4 & -7 & 1.1 $\pm$ 0.4$^*$ & -5 & 0.8 $\pm$ 0.4$^*$ & -3 & 1.3 $\pm$ 0.4     & -6 &  $<$ 0.6      & -   \\
2.208 & NaI        & $<$ 0.6       &  - &-0.7 $\pm$ 0.3$^*$ & +3 &-1.0 $\pm$ 0.4$^*$ & +4 &-0.8 $\pm$ 0.4$^*$ & +4 &-1.1 $\pm$ 0.3 &+4.5 \\
2.293 & CO 2-0     & 0.6 $\pm$ 0.2 & -3 &-1.3 $\pm$ 0.5$^*$ & +6 &-0.8 $\pm$ 0.4$^*$ & +3 &$<$1.2             & -  &-1.9 $\pm$ 0.3 & +8  \\
2.323 & CO 3-1     & 0.9 $\pm$ 0.2 & -5 &     -             & -  &     -             & -  &    -              & -  &-1.8 $\pm$ 0.3 & +7  \\
\enddata

\tablecomments{The same as in Table~\ref{linesxztau:tab}. Upper
limits are given as 3$\sigma$ values.}

\end{deluxetable}

\begin{deluxetable}{cccccc}
\tabletypesize{\footnotesize} \tablecaption{Line emission fluxes
of V1143 Ori. \label{linesv1143ori:tab}} \tablewidth{0pt}
\tablehead{ \colhead{$\lambda_{vac}$} & \colhead{Ident.} &
\colhead{F $\pm$ $\Delta$F} & \colhead{EW} & \colhead{F $\pm$
$\Delta$F} & \colhead{EW}  } \startdata
($\mu$m) &   & \multicolumn{2}{c}{07/Mar/17} & \multicolumn{2}{c}{07/Oct/15} \\
\tableline
1.094    &  Pa$\gamma$ &   $<$ 0.3              & -   & 0.4 $\pm$ 0.2$^*$  & -13  \\
1.282    &  Pa$\beta$  &   $<$ 0.3              & -   &   $<$ 0.3          & -    \\
2.166    &  Br$\gamma$ &   $<$ 0.2              &  -  &   $<$ 0.8          &      \\
2.208    &  NaI        &   $<$ 0.2              &  -  & -1.5 $\pm$ 0.3     & +1   \\
2.293    &  CO 2-0     &  -0.3 $\pm$ 0.1        & +33 &   $<$ 0.4          & -    \\
2.323    &  CO 3-1     &  -0.3 $\pm$ 0.1        & +29 &    -               & -    \\
\enddata

\tablecomments{The same as in Table~\ref{linesxztau:tab}. Upper
limits are given as 3$\sigma$ values.}

\end{deluxetable}

\begin{deluxetable}{cccccccccccc}
\tabletypesize{\scriptsize} \tablecaption{Line emission fluxes of
PV Cep. \label{linespvcep:tab}} \tablewidth{0pt} \tablehead{
\colhead{$\lambda_{vac}$} & \colhead{Ident.} & \colhead{F $\pm$
$\Delta$F} & \colhead{EW} & \colhead{F $\pm$ $\Delta$F} &
\colhead{EW} & \colhead{F $\pm$ $\Delta$F} & \colhead{EW} &
\colhead{F $\pm$ $\Delta$F} & \colhead{EW} & \colhead{F $\pm$
$\Delta$F} & \colhead{EW}} \startdata
($\mu$m) &   & \multicolumn{2}{c}{07/May/13} & \multicolumn{2}{c}{07/Oct/15} & \multicolumn{2}{c}{07/Oct/27} & \multicolumn{2}{c}{08/Apr/01} & \multicolumn{2}{c}{08/Apr/12} \\
\tableline
0.836    &  Pa22       & 1.0 $\pm$ 0.1     &  -41 &    -             &  -   &      -           & -   &         -      &  - &        -          & -  \\
0.850    &  Pa16       & 3.5 $\pm$ 0.4     & -138 & 2.8 $\pm$ 0.4    & -176 &    $<$ 4         & -   & 4.0 $\pm$ 0.4  &-53 & 1.0 $\pm$ 0.4$^*$ &-14 \\
0.867    &  Pa13       & 1.5 $\pm$ 0.1     &  -68 & 2.4 $\pm$ 0.1    & -136 & 3.2 $\pm$ 0.1    &-254 & 2.5 $\pm$ 0.4  &-32 & 2.0 $\pm$ 0.4     &-28 \\
0.923    &  Pa9        & 0.9 $\pm$ 0.3     &  -36 &      -           &  -   &      -           &   - & 2.0 $\pm$ 0.3  &-21 & 1.6 $\pm$ 0.3     &-20 \\
1.005    &  Pa$\delta$ & 1.1 $\pm$ 0.3     &  -33 & 0.6 $\pm$ 0.3$^*$& -18 & 0.7 $\pm$ 0.3$^*$ & -18 & 1.6 $\pm$ 0.3  &-13 & 1.2 $\pm$ 0.3     &-11 \\
1.070    &   ?         & 1.0 $\pm$ 0.4$^*$ &  -23 & 0.8 $\pm$ 0.4$^*$& -15 & 1.1 $\pm$ 0.4$^*$ & -20 & 2.2 $\pm$ 0.4  &-21 & 1.5 $\pm$ 0.4     &-10 \\
1.094    &  Pa$\gamma$ & 1.2 $\pm$ 0.2     &  -25 & 0.9 $\pm$ 0.2    & -18 & 1.0 $\pm$ 0.2     & -17 & 3.2 $\pm$ 0.2  &-17 & 2.6 $\pm$ 0.2     &-16 \\
1.137    &  OI         & 0.8 $\pm$ 0.2     &  -16 & 0.8 $\pm$ 0.2    & -15 & 0.5 $\pm$ 0.2$^*$ & -7  & 1.7 $\pm$ 0.2  &-9  & 2.3 $\pm$ 0.2     &-12 \\
1.189    &  [PII]      &    -              &  -   & 0.8 $\pm$ 0.2    & -11 &      -            &   - &    -           & -  &     -             &  - \\
1.205    &   ?         & 1.3 $\pm$ 0.4     &  -19 & 0.8 $\pm$ 0.4$^*$& -9  &      -            &   - &    -           & -  &     -             &  - \\
1.264    &   ?         & 1.0 $\pm$ 0.4$^*$ &  -13 & 0.8 $\pm$ 0.4$^*$& -8  & 0.6 $\pm$ 0.2     & -5  & 0.9 $\pm$ 0.3  & -3 & 0.8$\pm$ 0.2      & -3 \\
1.282    &  Pa$\beta$  & 3.8 $\pm$ 0.2     &  -48 & 3.7 $\pm$ 0.2    & -40 & 3.5 $\pm$ 0.2     & -28 & 5.3 $\pm$ 0.2  &-16 & 6.1 $\pm$ 0.2     &-20 \\
1.504    &  Br23+MgI   & 1.7 $\pm$ 0.3     &   -9 & 1.9 $\pm$ 0.3    & -7  & 2.5 $\pm$ 0.3     & -10 & 5.6 $\pm$ 0.3  &-8  & 3.8 $\pm$ 0.3     &-6  \\
1.570    &  Br15       & 0.8 $\pm$ 0.3$^*$ &   -4 &     -            & -   &      -            &   - &    -           & -  &     -             & -  \\
1.588    &  Br14       & 1.4 $\pm$ 0.2     &   -7 & 2.4 $\pm$ 0.2    & -8  & 2.1 $\pm$ 0.2     & -7  & 5.2 $\pm$ 0.2  &-7  & 3.6 $\pm$ 0.2     &-5  \\
1.611    &  Br13       & 2.0 $\pm$ 0.4     &   -9 & 2.3 $\pm$ 0.4    & -7  & 3.3 $\pm$ 0.4     & -11 & 5.5 $\pm$ 0.4  &-7  & 4.2 $\pm$ 0.4     &-6  \\
1.641    &  Br12       & 3.6 $\pm$ 0.4     &  -16 & 3.8 $\pm$ 0.4    & -11 & 2.9 $\pm$ 0.4     & -9  & 5.0 $\pm$ 0.4  &-6  & 5.0 $\pm$ 0.4     &-7  \\
1.681    &  Br11       & 2.6 $\pm$ 0.3     &  -11 & 4.9 $\pm$ 0.3    & -14 & 2.9 $\pm$ 0.3     & -9  & 7.1 $\pm$ 0.3  &-9  & 7.4 $\pm$ 0.3     &-9  \\
1.737    &  Br10       & 3.5 $\pm$ 0.3     &  -13 & 4.2 $\pm$ 0.3    & -20 & 4.6 $\pm$ 0.3     & -12 &13.8 $\pm$ 0.3  &-18 & 8.4 $\pm$ 0.3     &-11 \\
1.945    &  Br$\delta$ & 5.4 $\pm$ 0.3     &  -15 &     -            &  -  &      -            & -   &    -           & -  &     -             & -  \\
2.166    &  Br$\gamma$ & 8.5 $\pm$ 0.4     &  -17 & 7.8 $\pm$ 0.4    & -11 & 7.3 $\pm$ 0.4     & -11 & 8.3 $\pm$ 0.4  &-7  & 9.6 $\pm$ 0.4     &-8  \\
2.208    &  NaI        & 2.6 $\pm$ 0.3     &  -5  & 5.3 $\pm$ 0.3    & -7  & 2.1 $\pm$ 0.3     & -3  & 5.1 $\pm$ 0.3  &-4  & 3.9 $\pm$ 0.3     &-3  \\
2.293    &  CO 2-0     & 9.7 $\pm$ 0.7     &  -20 & 11.4 $\pm$ 0.7   & -13 & 8.8 $\pm$ 0.7     & -10 &18.0 $\pm$ 0.7  &-13 &11.6 $\pm$ 0.7     &-9  \\
2.323    &  CO 3-1     & 12.8 $\pm$ 0.7    &  -20 & 10.9 $\pm$ 0.7   & -12 & 10.3 $\pm$ 0.7    & -10 &20.2 $\pm$ 0.7  &-15 &11.7 $\pm$ 0.7     &-8  \\
\tableline
 &             & \multicolumn{2}{c}{08/Jun/10} & \multicolumn{2}{c}{08/Jun/18} &  &  &  \\
\tableline
1.005 & Pa$\delta$ & 0.7 $\pm$ 0.2      & -42  &   -              &  -   \\
1.094 & Pa$\gamma$ & 0.5 $\pm$ 0.2$^*$  & -20 & 0.5 $\pm$ 0.2$^*$ & -19  \\
1.264 &   ?        & 0.6 $\pm$ 0.2      & -12 & 0.6 $\pm$ 0.2     & -17  \\
1.282 & Pa$\beta$  & 1.4 $\pm$ 0.2      & -22 & 1.2 $\pm$ 0.2     & -19  \\
2.166 & Br$\gamma$ & 3.5 $\pm$ 0.4      & -6  & 3.3 $\pm$ 0.6     & -8   \\
2.208 &  NaI       & $<$ 1.3            &  -  & $<$ 1.4           & -    \\
2.293 &  CO 2-0    & 1.3 $\pm$ 0.4      & -2  & 2.6 $\pm$ 0.6     & -5   \\
2.323 &  CO 3-1    & 1.8 $\pm$ 0.5      & -3 & 1.8  $\pm$ 0.6     & -3   \\
\tableline

\enddata

\tablecomments{The same as in Table~\ref{linesxztau:tab}.}

\end{deluxetable}

\begin{deluxetable}{lccccccc}
\tabletypesize{\footnotesize} \tablecaption{EXor line flux
variability in the near IR. \label{linevar:tab}} \tablewidth{0pt}
\tablehead{ \colhead{Source} & \colhead{Date} &
\colhead{Pa$\gamma$} & \colhead{Pa$\beta$} & \colhead{Br$\gamma$}
& \colhead{CO (2-0)} & \colhead{Status$^a$} & \colhead{Ref}}
\startdata
   &    & \multicolumn{4}{c}{(10$^{-13}$ erg s$^{-1}$ cm$^{-2}$)} &  & \\
\tableline
XZ Tau      & 83 Nov           &  -           &  -             &  2.1 $\pm$ 0.5    &  -             & H ? &  2   \\
            & 86 May - 87 Jan  &  -           &  -             & 0.42 $\pm$ 0.07   &  -             & H ? &  3   \\
            & 88 Nov - 89 Jan  &  -           & 2.4 $\pm$ 0.4  &  0.9 $\pm$ 0.1    &  -             &     &  4   \\
            & {\bf 07 Mar}     & {\bf $<$ 0.6}     & {\bf 6 $\pm$ 1 }   & {\bf 2.4 $\pm$ 0.2 }  & {\bf -2.5 $\pm$ 0.6}   &     & {\bf 1  }\\
            & {\bf 07 Oct}     & {\bf 3.5 $\pm$ 1} & {\bf 6 $\pm$ 2 }   & {\bf 2.5 $\pm$ 0.2 }  & {\bf -2.7 $\pm$ 0.6}   &     & {\bf 1  }\\
            & {\bf 08 Mar}     & {\bf $<$ 1.2}     & {\bf 6 $\pm$ 1 }   & {\bf 1.6 $\pm$ 0.4 }  & {\bf -5.3 $\pm$ 0.4}   &     & {\bf 1  }\\
\tableline
UZ Tau E    & {\bf 07 Mar}     & {\bf 9 $\pm$ 1}   &{\bf 10.9 $\pm$ 0.8}& {\bf 4.5 $\pm$ 0.8 }  & {\bf -4.2 $\pm$ 0.8}   &     & {\bf 1  }\\
            & {\bf 07 Oct}     & {\bf 9 $\pm$ 1}   &{\bf 10.4 $\pm$ 0.8}& {\bf 5.2 $\pm$ 0.8 }  & {\bf -3.7 $\pm$ 0.8}   &     & {\bf 1  }\\
            & {\bf 08 Mar}     & {\bf 6 $\pm$ 1}   & {\bf 10 $\pm$ 1 }  & {\bf 4 $\pm$ 1}       & {\bf $<$ 2.0}          &     & {\bf 1  }\\
\tableline
VY Tau      & {\bf 07 Mar}     & {\bf $<$ 1.8}     &{\bf 1.9 $\pm$ 0.4} & {\bf $<$ 0.6}         & {\bf $<$ 1.0}          &     & {\bf 1  }\\
            & {\bf 07 Oct}     & {\bf 3 $\pm$ 1}   &{\bf 1.7 $\pm$ 0.4} & {\bf $<$ 0.9}         & {\bf -2.0 $\pm$ 0.5}   &     & {\bf 1  }\\
            & {\bf 08 Mar}     &{\bf 3.1 $\pm$ 0.9}&{\bf 1.9 $\pm$ 0.4 }& {\bf 0.7 $\pm$ 0.2}   & {\bf $<$ 0.9}          &     & {\bf 1  }\\
\tableline
DR Tau      & 88 Nov - 89 Jan  &  -           & 12 $\pm$ 1.2   &  3.2 $\pm$ 0.5    &  -             & I ? &  4   \\
            & {\bf 07 Mar}     & {\bf 26 $\pm$ 0.8} &{\bf 37 $\pm$ 0.8}  & {\bf 10 $\pm$ 1 }     & {\bf 3 $\pm$ 1}        &     & {\bf 1  }\\
            & {\bf 07 Oct}     & {\bf 18 $\pm$ 2}   &{\bf 24 $\pm$ 2}    & {\bf 6.2 $\pm$ 0.6}   & {\bf 2.0 $\pm$ 0.6}    &     & {\bf 1  }\\
            & {\bf 07 Nov}     & {\bf 25 $\pm$ 2}   &{\bf 32 $\pm$ 2}    & {\bf 7.5 $\pm$ 0.6}   & {\bf -7.1 $\pm$ 0.6}   &     & {\bf 1  }\\
            & {\bf 08 Mar}     &{\bf 20.0 $\pm$ 0.5}&{\bf 30.0 $\pm$ 0.5}& {\bf 7 $\pm$ 1}       & {\bf -9.4 $\pm$ 0.6}   &     & {\bf 1  }\\
\tableline
V1118 Ori   & {\bf 05 Sep}     & {\bf 2.5 $\pm$ 0.2}&{\bf 4.1 $\pm$ 0.1 }& {\bf 1.2 $\pm$ 0.2}   & {\bf 1.2 $\pm$ 0.4}    &  H  & {\bf 5  }\\
            & {\bf 06 Sep}     & {\bf $<$ 0.4}      &{\bf $<$ 0.6}       & {\bf $<$ 0.2}         & {\bf -}                &     & {\bf 6  }\\
\tableline
NY Ori      & {\bf 07 Mar}     & {\bf 2.5 $\pm$ 0.5}&{\bf 3.2 $\pm$ 0.5} & {\bf 1.5 $\pm$ 0.4}   & {\bf 0.6 $\pm$ 0.2}    &     & {\bf 1  }\\
            & {\bf 07 Oct}     & {\bf 3 $\pm$ 0.5}  &{\bf 4.0 $\pm$ 0.5} & {\bf $<$ 1.2}         & {\bf $<$ 1.5}          &     & {\bf 1  }\\
            & {\bf 07 Oct}     & {\bf 1.5 $\pm$ 0.5}&{\bf 2.5 $\pm$ 0.5} & {\bf $<$ 1.2}         & {\bf $<$ 1.2}          &     & {\bf 1  }\\
            & {\bf 07 Oct}     & {\bf $<$ 1.5}      &{\bf 2.2 $\pm$ 0.5} & {\bf 1.3 $\pm$ 0.4}   & {\bf $<$ 1.2}          &     & {\bf 1  }\\
            & {\bf 08 Mar}     &{\bf 1.7 $\pm$ 0.5} &{\bf 3.3 $\pm$ 0.5} & {\bf $<$ 0.6}         & {\bf -1.9 $\pm$ 0.3}   &     & {\bf 1  }\\
\tableline
V1143 Ori   & {\bf 07 Mar}     & {\bf $<$ 0.3}      &{\bf $<$ 0.3}       & {\bf $<$ 0.2}         & {\bf -0.3 $\pm$ 0.1}   &     & {\bf 1  }\\
            & {\bf 07 Oct}     & {\bf $<$ 0.6}      &{\bf $<$ 0.3}       & {\bf $<$ 0.8}         & {\bf $<$ 0.4}          &     & {\bf 1  }\\
\tableline
PV Cep      & 86 Jul           &  -           &  -             & 0.44 $\pm$ 0.05   &  -             &  &  3   \\
            & 86 Oct           &  -           &  -             & 0.80 $\pm$ 0.12   &  -             &  &  3   \\
            & {\bf 07 Mar}     &{\bf 1.2 $\pm$ 0.2} &{\bf 3.8 $\pm$ 0.2} & {\bf 8.5 $\pm$ 0.4 }  & {\bf 9.7 $\pm$ 0.7}    &     & {\bf 1  }\\
            & {\bf 07 Oct}     &{\bf 0.9 $\pm$ 0.2} &{\bf 3.7 $\pm$ 0.2} & {\bf 7.8 $\pm$ 0.4}   & {\bf 11.4 $\pm$ 0.7}   &     & {\bf 1  }\\
            & {\bf 07 Oct}     &{\bf 1.0 $\pm$ 0.2} &{\bf 3.5 $\pm$ 0.2} & {\bf 7.3 $\pm$ 0.4}   & {\bf 8.8 $\pm$ 0.7}    &     & {\bf 1  }\\
            & {\bf 08 Apr}     &{\bf 3.2 $\pm$ 0.2} &{\bf 5.3 $\pm$ 0.2} & {\bf 8.3 $\pm$ 0.4}   & {\bf 18 $\pm$ 0.7}     &     & {\bf 1  }\\
            & {\bf 08 Apr}     &{\bf 2.6 $\pm$ 0.2} &{\bf 6.1 $\pm$ 0.2} & {\bf 9.6 $\pm$ 0.4}   & {\bf 11.6 $\pm$ 0.7}   &     & {\bf 1  }\\
\tableline
            &                  &                    &                    &                       &                        &     &      \\
            &                  &                    &                    &                       &                        &     &      \\
            &                  &                    &                    &                       &                        &     &      \\
%
   &    & \multicolumn{4}{c}{Equivalent Width (in \AA)} & \\
\tableline
XZ Tau      & 02 Nov           & -1.0         &  -   &   -  &  -       &  &  7   \\
            & {\bf 07 Mar}     & {\bf -}      & {\bf -10}   & {\bf -4} & {\bf +5}    &     & {\bf 1  }\\
            & {\bf 07 Oct}     & {\bf -4}     & {\bf -7}    & {\bf -4} & {\bf +5}    &     & {\bf 1  }\\
            & {\bf 08 Mar}     & {\bf -}      & {\bf -7}    & {\bf -2} & {\bf +8}    &     & {\bf 1  }\\
\tableline
UZ Tau E    & 02 Nov           & -4.0         &  -   &      &  -       & H ? &  7   \\
            & {\bf 07 Mar}     & {\bf -11}    & {\bf -13}   & {\bf -6} & {\bf +6}    &     & {\bf 1  }\\
            & {\bf 07 Oct}     & {\bf -7}     & {\bf -10}   & {\bf -9} & {\bf +7}    &     & {\bf 1  }\\
            & {\bf 08 Mar}     & {\bf -4}     & {\bf -8}    & {\bf -6} & {\bf -}     &     & {\bf 1  }\\
\tableline
VY Tau      & {\bf 07 Mar}     & {\bf -}      & {\bf -6}    & {\bf -}  & {\bf -}     &     & {\bf 1  }\\
            & {\bf 07 Oct}     & {\bf -11}    & {\bf -6}    & {\bf -}  & {\bf +19}   &     & {\bf 1  }\\
            & {\bf 08 Mar}     & {\bf -9}     & {\bf -6}    & {\bf -6} & {\bf -}     &     & {\bf 1  }\\
\tableline
DR Tau      & 98 Jan           &  -           &-21.2 & -7.3 &  -      & I &  8   \\
            & 02 Nov           & -12.3/-13.7  &  -   &   -  &  -      &   &  7   \\
            & {\bf 07 Mar}     & {\bf -17}    & {\bf -24}   & {\bf -8} & {\bf -3}    &     & {\bf 1  }\\
            & {\bf 07 Oct}     & {\bf -22}    & {\bf -26}   & {\bf -6} & {\bf -2}    &     & {\bf 1  }\\
            & {\bf 07 Nov}     & {\bf -21}    & {\bf -29}   & {\bf -7} & {\bf +7}    &     & {\bf 1  }\\
            & {\bf 08 Mar}     & {\bf -20}    & {\bf -31}   & {\bf -6} & {\bf +9}    &     & {\bf 1  }\\
\tableline
V1118 Ori   & {\bf 05 Sep}     & {\bf -20}    &{\bf -40}    & {\bf -30}& {\bf -30}   &     & {\bf 5  }\\
            & {\bf 06 Sep}     & {\bf -}      &{\bf -}      & {\bf -}  & {\bf -}     &     & {\bf 6  }\\
\tableline
NY Ori      & {\bf 07 Mar}     & {\bf -8}     & {\bf -11}   & {\bf -7} & {\bf -3}    &     & {\bf 1  }\\
            & {\bf 07 Oct}     & {\bf -9}     & {\bf -13}   & {\bf -5} & {\bf -}     &     & {\bf 1  }\\
            & {\bf 07 Oct}     & {\bf -5}     & {\bf -8}    & {\bf -3} & {\bf -}     &     & {\bf 1  }\\
            & {\bf 07 Oct}     & {\bf -}      & {\bf -7}    & {\bf -6} & {\bf -}     &     & {\bf 1  }\\
            & {\bf 08 Mar}     & {\bf -5}     & {\bf -10}   & {\bf -}  & {\bf +8}    &     & {\bf 1  }\\
\tableline
V1143 Ori   & {\bf 07 Mar}     & {\bf -}      & {\bf -}     & {\bf -}  & {\bf +33}   &     & {\bf 1  }\\
            & {\bf 07 Oct}     & {\bf -}      & {\bf -}     & {\bf -}  & {\bf -}     &     & {\bf 1  }\\
\tableline
PV Cep      & 86               &  -           &  -   &  -   & $<$ -3.1&  &  3   \\
            & 94 Jun           &  -           &  -   &  -   &  -4.4   &  &  9   \\
            & 95 Jan           &  -           & -18.2& -2.4 &    -    &  & 10   \\
            & {\bf 07 Mar}     & {\bf -25}    & {\bf -48}   & {\bf -17} & {\bf -20}  &     & {\bf 1  }\\
            & {\bf 07 Oct}     & {\bf -18}    & {\bf -40}   & {\bf -11} & {\bf -13}  &     & {\bf 1  }\\
            & {\bf 07 Oct}     & {\bf -17}    & {\bf -28}   & {\bf -11} & {\bf -10}  &     & {\bf 1  }\\
            & {\bf 08 Apr}     & {\bf -17}    & {\bf -16}   & {\bf -7}  & {\bf -13}  &     & {\bf 1  }\\
            & {\bf 08 Apr}     & {\bf -16}    & {\bf -20}   & {\bf -8}  & {\bf -9}   &     & {\bf 1  }\\
\enddata

\tablecomments{References to the Table: {\bf (1) Present paper};
(2) Evans et al. 1987; (3) Carr 1990; (4) Giovanardi et al. 1991;
(5) Paper I; (6) Paper II; (7) Edwards et al. 2006; (8) Muzerolle,
Hartmann \& Calvet 1998; (9) Biscaya et al. 1997; (10) Greene \&
Lada 1996.}

\tablenotetext{a}{~~~~ H = high, I = intermediate, L = low, r =
rising, d = declining }
\end{deluxetable}


\begin{deluxetable}{lcccccccccc}
\tabletypesize{\footnotesize} \tablecaption{EXor wind parameters.
\label{models:tab}} \tablewidth{0pt} \tablehead{ \colhead{Source}&
\colhead{Date}      & \colhead{A$_V$}     & \colhead{R$_{out}$} &
\colhead{T} & \colhead{Pa$\beta$} & \colhead{Br$\gamma$}&
\colhead{Pa$\beta$/Br$\gamma$} & \colhead{Pa$\beta$/Br$\gamma$} &
\colhead{$\dot{M}_{wind}$}} \startdata
   &  & (mag) & (R$_{*})$  & (K) &
\multicolumn{2}{c}{(mod/obs)} & (mod) & (obs) & (10$^{-7}$ M$_{\sun}$ yr$^{-1}$) \\
\tableline
UZ Tau E    & 07 Mar 12  & 1.5 & 3.0 & 6000  & 1.2 & 0.9 & 4.4 & 3.3 & 0.1  \\
            & 07 Oct 11  & 1.5 & 3.0 & 6000  & 1.4 & 0.9 & 4.4 & 2.5 & 0.1  \\

DR Tau      & 07 Mar 13  & 1.9 & 3.0 & 7400  & 1.0 & 1.0 & 3.6 & 3.7 & 0.6  \\
            & 07 Oct 08  & 1.9 & 3.0 & 6400  & 1.0 & 1.0 & 3.8 & 3.9 & 0.3  \\
            & 07 Nov 03  & 1.9 & 3.0 & 6400  & 1.0 & 1.3 & 3.1 & 4.3 & 0.6  \\
            & 08 Mar 28  & 1.9 & 3.0 & 8000  & 1.0 & 1.0 & 4.1 & 4.3 & 0.4  \\

V1118 Ori   & 05 Sep 10  & 0   & 5.0 & 8000  & 1.1 & 1.1 & 3.6 & 3.4 & 0.4  \\

PV Cep      & 07 May 13  & 11  & 5.0 & 6000  & 1.0 & 0.7 & 0.6 & 2.3 & 9.7  \\
            & 07 Oct 15  & 13  & 7.0 & 6000  & 1.0 & 0.6 & 0.7 & 3.3 & 5.1  \\
            & 07 Oct 27  & 11  & 20  & 6000  & 1.0 & 0.4 & 1.3 & 2.5 & 2.0  \\
            & 08 Apr 01  & 9   & 4.5 & 6000  & 1.0 & 0.8 & 0.8 & 2.5 & 21   \\
            & 08 Apr 12  & 9   & 4.5 & 6000  & 1.0 & 0.6 & 0.9 & 2.5 & 6.7  \\
            & 08 Jun 10  &14.5 & 3.5 & 7000  & 1.0 & 1.0 & 0.4 & 3.5 & 22   \\
            & 08 Jun 18  &14.5 & 4.0 & 6000  & 1.0 & 0.9 & 0.4 & 3.2 & 8.2  \\
\enddata
\end{deluxetable}


\begin{deluxetable}{lcccccc}
\tabletypesize{\footnotesize} \tablecaption{EXor accretion
parameters.\label{modelsacc:tab}} \tablewidth{0pt} \tablehead{
\colhead{Source} & \colhead{Date} & \colhead{$\Delta$L$_{JHK}$} &
\colhead{A$_V$} & \colhead{L$_{acc}$(Pa$\beta$)} &
\colhead{L$_{acc}$(Br$\gamma$)} & \colhead{$\dot{M}_{acc}$} }
\startdata
   & & ($L_{\sun}$) & (mag) & ($L_{\sun}$) & ($L_{\sun}$)
   & (10$^{-7}$ M$_{\sun}$ yr$^{-1}$) \\
\tableline
UZ Tau E    & 07 Mar 12  &   -   & 1.5 & 0.5 & 1.0 & 1-3 \\
            & 07 Oct 11  & - 0.1 & 1.5 & 0.5 & 1.2 & 1-3 \\

DR Tau      & 07 Mar 13  &   -   & 1.9 & 2.3 & 2.9 & 7-9 \\
            & 07 Oct 08  & - 0.2 & 1.9 & 0.9 & 1.3 & 3-4 \\
            & 07 Nov 03  & - 0.1 & 1.9 & 2.0 & 2.0 &  6  \\
            & 08 Mar 28  & - 0.1 & 1.9 & 1.8 & 1.9 &  4  \\

V1118 Ori   & 05 Sep 10  &   -   & 0   & 1.5 & 2.5 & 4-7 \\
            & 06 Sep 23  & - 0.4 & 2.5 & - & - & - \\

PV Cep      & 07 May 13  &   -    & 11 & 41.7 & 182  & 25-100  \\
            & 07 Oct 15  & + 1.3  & 13 & 72.4 & 219  & 47-130  \\
            & 07 Oct 27  & + 2.4  & 11 & 38.0 & 158  & 23-95   \\
            & 08 Apr 01  & + 8.0  & 9  & 34.7 & 147  & 21-88   \\
            & 08 Apr 12  & + 7.6  & 9  & 40.7 & 170  & 24-100  \\
            & 08 Jun 10  & + 0.6  &14.5& 37.0 & 91.6 & 22-55   \\
            & 08 Jun 18  & - 0.9  &14.5& 31.5 & 81.3 & 19-49   \\
\enddata
\end{deluxetable}

\begin{deluxetable}{lccccc}
\tabletypesize{\footnotesize} \tablecaption{Parameters of the
candidate EXor's. \label{candidates:tab}} \tablewidth{0pt}
\tablehead{ \colhead{Source} & \colhead{Status} & \colhead{Near-IR
spect.} & \colhead{P$_{Cyg}$} & \colhead{B or R while} &
\colhead{Ref.} } \startdata
 &  &  &  & brightening  &  \\
\tableline
             &         &                                   &                  &          &     \\
SVS 13       & rising  & increasing CO em.                 & PC H$\alpha$, HH + CO & blueing  & 1,2,3 \\
             &         & HI, NaI, CO, H$_2$, [FeII]        &                  &          & 4,5,6      \\
             &         & CO (but no EW) variability        & narrow line      &          & 7   \\
L1415 IRS    & rising  & no IR spectrum                    & PC H$\alpha$, HH &          & 8   \\
V1647 Ori    &outburst & Br$\gamma$, NaI, CO               & PC H$\alpha$     & blueing  & 9,10 \\
             &outburst & recomb., ionic, CO                & PC Paschen, HeI  &          & 11  \\
             & fading  & recomb., ionic, CO                & PC absent        &          & 12  \\
             & fading  & CO rovibrational                  &    --            &          & 13  \\
             &quiescent& line em. steady decrease          & PC absent        &          & 14  \\
             &quiescent& CO abs., HI,He fading             &    --            &          & 15  \\
ISO Cha I 192& rising  & recomb., ionic, CO                & CO outflow       & blueing  & 16  \\
OO Ser       & rising  & featureless, rising cont.         &    --            &          & 17  \\
             & fading  & faint CO bandhead abs.            &    --            &          & 18  \\
             & fading  & featureless, rising cont.         &    --            &          & 19  \\
EC 53        & fading  & no IR spectrum                    &    --            & blueing           & 17 \\
             & variab. & no IR spectrum                    &    --            & no color var.     & 18 \\
             &         & abs. features                     &    --            &                   & 20 \\
GM Cep       & rising  & no IR spectrum                    & PC H$\alpha$     & random color var. & 21 \\
\enddata

\tablecomments{(1) Eisl\"{o}ffel et al. 1991; (2) Liseau,
Lorenzetti \& Molinari 1992; (3) Carr 1990; (4) Davis et al. 2006;
(5) Takami et al. 2006 (and references therein); (6) Davis et al.
2002; (7) Biscaya et al. 1997; (8) Stecklum, Melnikov \& Meusinger
2007; (9) Reipurth \& Aspin 2004; (10) Walter et al. 2004; (11)
Vacca, Cushing \& Simon 2004; (12) Gibb et al. 2006; (13) Brittain
et al. 2007; (14) Acosta-Pulido et al. 2007; (15) Aspin, Beck \&
Reipurth 2008; (16) G\'{o}mez \& Mardones 2003; (17) Hodapp et al.
1996; (18) Hodapp 1999; (19) K\'{o}sp\'{a}l et al. 2007; (20)
Doppmann et al. 2005; (21) Sicilia-Aguilar et al. 2007.}

\tablenotetext{a}{~~~~ PC = P$_{Cyg}$ profile; IPC = inverse
P$_{Cyg}$ profile; EW = Equivalent Width. }

\end{deluxetable}


\vspace{2cm}

\begin{thebibliography}{}


\bibitem{} Acosta-Pulido, J.A., Kun, M., \'{A}brah\'{a}m, P. et al. 2007, AJ,133, 2020
\bibitem{} Antoniucci, S., Nisini, B., Giannini, T. \& Lorenzetti, D. 2007 A\&A 479, 503
\bibitem{} Aspin, C., Beck, T.L., \& Reipurth, B. 2008 AJ 135, 423
\bibitem{} Beck, T., McGregor, P.J., Takami, M. \& Tae-Soo Pyo 2008 ApJ 676, 472
\bibitem{} Basri, G. \& Bertout, C. 1989 ApJ 341, 340
\bibitem{} Beristain, G., Edwards, S. \& Kwan, J. 1998 ApJ 499, 828
\bibitem{} Biscaya, A.M., Rieke, G.H., Narayanan, Gopal, Luhman, K.L. \& Young E.T. 1997 ApJ 491, 359
\bibitem{} Bonnell, L. \& Bastien, P. 1992 ApJ 401, L31
\bibitem{} Breger, M., Gehrz, R.D. \& Hackwell, J.A. 1981 ApJ 248, 963
\bibitem{} Carr, J.S. 1989 ApJ 345, 522
\bibitem{} Carr, J.S. 1990 AJ 100, 1244
\bibitem{} Clarke, C., Lin, D. \& Pringle, J. 1990 MNRAS 242, 439
\bibitem{} Coffey, D., Downes, T.P. \& Ray, T.P. 2004 A\&A 419, 593
\bibitem{} Cohen, M., Kuhi, L.V. 1979 ApJS, 41, 743
\bibitem{} Cohen, M., Kuhi, L.V., Harlan, E.A. \& Spinrad, H. 1981 ApJ 245, 920
\bibitem{} Davis, C.J., Nisini, B., Takami, M. et al. 2006 ApJ 639, 969
\bibitem{} Davis, C.J., Stern, L., Ray, T.P. \& Chrysostomou, A. 2002 A\&A 382, 1021
\bibitem{} Doppmann, G.W., Greene, T.P., Covey, K.R. \& Lada, C.J. 2005 AJ 130, 1145
\bibitem{} Elias, J.H. 1978 ApJ 224, 857
\bibitem{} Edwards, S., Fisher, W., Hillenbrand, L. \& Kwan, J. 2006 ApJ 646, 319
\bibitem{} Edwards, S., Fisher, W., Kwan, J., Hillenbrand, L. \& Dupree, A.K. 2003 ApJ 599, L41
\bibitem{} Eisl\"{o}ffel, J., G\"{u}nter, E., Hessman, F.V. et al. 1991 ApJL 383, L19
\bibitem{} Evans II, N.J., Levreault, R.M., Beckwith, S. \& Strutskie, M. 1987 ApJ 320, 364
\bibitem{} Folha, D.F.M. \& Emerson, J.P. 1999 A\&A 352, 517
\bibitem{} Folha, D.F.M. \& Emerson, J.P. 2001 A\&A 365, 90
\bibitem{} Gasparian, L.G., Melkonian, A.S., Ohanian, G.B. \& Parsamian, E.S. 1990 {\it Flare Stars in Star Clusters,
Associations, and the Solar Vicinity} Proc. 137th Symposium of the
International Astronomical Union, Byurakan [Armenia], U.S.S.R.,
October 23-27, 1989. Editors, L.V. Mirzoyan, B.R. Pettersen, \&
M.K. Tsvetkov; Publ. Kluwer Academic, Dordrecht, The Netherlands,
Boston, MA.
\bibitem{} Gibb, E.L., Rettig, T.W., Brittain, S.D. et al. 2006 ApJ 641, 383
\bibitem{} Giovanardi. C., Gennari, S., Natta, A. \& Stanga, R. 1991 ApJ 367, 173
\bibitem{} G\'{o}mez, M. \& Mardones, D. 2003 AJ 125, 2134
\bibitem{} Greene, T.P. \& Lada, C.J. 1996 AJ 112, 2184
\bibitem{} Gullbring, E., Hartmann, L., Bric\~{c}eno, C. \& Calvet, N. 1998 ApJ 492, 323
\bibitem{} Gyul'budagyan, A.L. \& Magakyan, T. Yu. 1977 Pis'ma Astron. Zh. 3, 162
\bibitem{} Hamann, F., Simon, M., Carr, J.S. \& Prato, L. 1994 ApJ 436, 292
\bibitem{} Hartigan, P., Edwards, S., \& Ghandour, L. 1995 ApJ 452, 736
\bibitem{} Hartmann, L., Hewett, R. \& Calvet, N. 1994 ApJ 426, 669
\bibitem{} Hartmann, L., Hinkle, K. \& Calvet, N. 2004 ApJ 609, 906
\bibitem{} Hartmann, L. \& Kenyon, S. 1985 ApJ 299, 462
\bibitem{} Herbig, G.H. 1989 Proc. of the ESO Workshop on {\it Low Mass Star Formation and
Pre-Main Sequence Objects}, ed. B. Reipurth, p.233
\bibitem{} Herbig, G.H. 1990 ApJ 360, 639
\bibitem{} Herbig, G.H. 2007 AJ 133, 2679
\bibitem{} Herbig, G.H. 2008 AJ 135, 637
\bibitem{} Herbig, G.H. \& Bell, K.R. 1988 Lick Obs. Bulletin No.1111
\bibitem{} Herbig, G.H., Aspin, C., Gilmore, A.C. et al. 2001 PASP 113, 1547
\bibitem{} Hodapp, K.W. 1999 AJ, 118, 1338
\bibitem{} Hodapp, K.W., Hora, J.L., Rayner, J.T. et al. 1996 ApJ 468, 861
\bibitem{} Jensen, E.L.N., Dhital, S., Stassun, K.G. et al. 2007 AJ 134, 241
\bibitem{} Kenyon, S.J., Hartmann, L., Hewett, R. et al. 1994 AJ 107, 2153
\bibitem{} K\"{o}hler, R., Petr-Gotzens, M.G., McCaughrean, M.J. et al. 2006 A\&A 458, 461
\bibitem{} K\"{o}nigl, A. \& Pudritz, R.E. 2000 Protostars and Planets IV - University of Arizona Press;
eds V.Mannings, A.P.Boss, S.S Russell, p.759
\bibitem{} K\'{o}sp\'{a}l, \'{A}., \'{A}brah\'{a}m, P. , Acosta-Pulido, J. et al. 2007 A\&A, 470, 211
\bibitem{} Kraus, S., Hofmann, K.-H., Benisty, M. et al. 2008 A\&A, in press (arXiv:0807.1119)
\bibitem{} Lan\c{c}on, A., \& Rocca-Volmerange, B. 1992 A\&A Supp. Ser. 96, 593
\bibitem{} Liseau, R., Lorenzetti, D. \& Molinari, S. 1992 A\&A 253, 119
\bibitem{} Lorenzetti, D., Giannini, T., Calzoletti, L. et al. 2006 A\&A 453, 579 (Paper I)
\bibitem{} Lorenzetti, D., Giannini, T., Larionov, V.M. et al. 2007 ApJ 665, 1182 (Paper II)
\bibitem{} Malbet, F., Benisty, M., de Wit, W.-J. et al. 2007 A\&A 464, 43
\bibitem{} Martin, S.C. 1997 ApJ 478, L33
\bibitem{} McGregor. P.J., Hyland, A.R. \& Hillier, D.J. 1988 ApJ 324, 1071
\bibitem{} Menten, K.M., Reid, M.J., T., Forbrich, J. \& Brunthaler, A. 2007 A\&A 474, 515
\bibitem{} Meyer, M.R., Calvet, N. \& Hillenbrand, L.A. 1997 AJ 114, 288
\bibitem{} Muzerolle, J., Calvet, N. \& Hartmann, L. 2001 ApJ 550, 944
\bibitem{} Muzerolle, J., Hartmann, L. \& Calvet, N. 1998 AJ 116, 2965
\bibitem{} Najita, J., Carr, J.S., Glassgold, A.E., Shu, F.H. \& Tokunaga, A.T. 1996
ApJ 462, 919
\bibitem{} Nisini, B., Antoniucci, S. \& Giannini, T. 2004 A\&A 421, 187
\bibitem{} Reipurth, B. \& Aspin, C. 2004 ApJ 606, L119
\bibitem{} Reipurth, B., Guimar\~{a}es, M.M., Connelly, M.S. \& Bally, J. 2007, AJ 134, 2272
\bibitem{} Sato, S., Nagata, T., Tanaka, M. \& Yamamoto, T. 1990 ApJ 359, 192
\bibitem{} Rieke, G.H. \& Lebofsky, M.J. 1985 ApJ 288, 618
\bibitem{} Scoville, N.Z., Krotkov, R. \& Wang, D. 1980 ApJ 240, 929
\bibitem{} Sicilia-Aguilar, A., Mer\'{\i}n, B., T., Hormuth, F. \& \'{A}brah\'{a}m, P. 2008 ApJ 673, 382
\bibitem{} Shu, F.H., Najita, J.R., Ostriker, E. et al. 1994 ApJ 429, 781
\bibitem{} Shu, F.H., Najita, J.R., Shang, -H. \& Li, Z.-Y. 2000 Protostars and Planets IV - University of Arizona Press;
eds V.Mannings, A.P.Boss, S.S Russell, p.789
\bibitem{} Stecklum, B., Melnikov, S.Y., \& Meusinger, H. 2007 A\&A 463, 621
\bibitem{} Takami, M., Chrysostomou, A., Ray, T.P. et al. 2006 ApJ 641, 357
\bibitem{} Tatulli, E., Malbet, F., Menard, F. et al. 2008 A\&A, in press (arXiv:0806.4937)
\bibitem{} Vacca, W.D., Cushing, M.C. \& Simon, T. 2004 ApJ 609, L29
\bibitem{} van Citters, G.W. \& Smith, R.G. 1989 AJ 98, 1328
\bibitem{} Walter, F.M., Stringfellow, G.S., Sherry, W.H., \& Field-Pollatou, A. 2004 ApJ 128, 1872
\bibitem{} Whelan, E.T., Ray, T.P. \& Davis, C.J. 2004 A\&A 417, 247

\end{thebibliography}
\end{document}